\documentclass[sigconf]{acmart}

\AtBeginDocument{%
  }

\copyrightyear{2025} 
\acmYear{2025} 
\setcopyright{acmlicensed}\acmConference[SIGIR '25]{Proceedings of the 48th International ACM SIGIR Conference on Research and Development in Information Retrieval}{July 13--18, 2025}{Padua, Italy}
\acmBooktitle{Proceedings of the 48th International ACM SIGIR Conference on Research and Development in Information Retrieval (SIGIR '25), July 13--18, 2025, Padua, Italy}
\acmDOI{10.1145/3726302.3729956}
\acmISBN{979-8-4007-1592-1/2025/07}

\usepackage[ruled]{algorithm2e} 
\usepackage{multirow}
\usepackage{enumitem}

\begin{document}

\title{DLF: Enhancing Explicit-Implicit Interaction via Dynamic Low-Order-Aware Fusion for CTR Prediction}



\author{Kefan Wang}
\affiliation{%
  \institution{University of Science and Technology of China}
  \city{Hefei}
  \state{Anhui}
  \country{China}}
\email{wangkefan@mail.ustc.edu.cn}

\author{Hao Wang}
\affiliation{%
  \institution{University of Science and Technology of China}
  \city{Hefei}
  \state{Anhui}
  \country{China}}
\email{wanghao3@ustc.edu.cn}
\authornote{Corresponding author.}

\author{Wei Guo}
\affiliation{%
  \institution{Huawei Noah’s Ark Lab}
  \city{Singapore}
  \country{Singapore}}
\email{guowei67@huawei.com}

\author{Yong Liu}
\affiliation{%
  \institution{Huawei Noah’s Ark Lab}
  \city{Singapore}
  \country{Singapore}}
\email{liu.yong6@huawei.com}

\author{Jianghao Lin}
\affiliation{%
  \institution{Shanghai Jiao Tong University}
  \city{Shanghai}
  \country{China}}
\email{chiangel@sjtu.edu.cn}

\author{Defu Lian}
\affiliation{%
  \institution{University of Science and Technology of China}
  \city{Hefei}
  \state{Anhui}
  \country{China}}
\email{liandefu@ustc.edu.cn}

\author{Enhong Chen}
\affiliation{%
  \institution{University of Science and Technology of China}
  \city{Hefei}
  \state{Anhui}
  \country{China}}
\email{cheneh@ustc.edu.cn}

\renewcommand{\shortauthors}{Kefan Wang et al.}

\begin{abstract}
Click-through rate (CTR) prediction is a critical task in online advertising and recommender systems, relying on effective modeling of feature interactions. Explicit interactions capture predefined relationships, such as inner products, but often suffer from data sparsity, while implicit interactions excel at learning complex patterns through non-linear transformations but lack inductive biases for efficient low-order modeling. Existing two-stream architectures integrate these paradigms but face challenges such as limited information sharing, gradient imbalance, and difficulty preserving low-order signals in sparse CTR data.
We propose a novel framework, \textbf{\underline{D}ynamic \underline{L}ow-Order-Aware \underline{F}usion} (\textbf{DLF}), which addresses these limitations through two key components: a Residual-Aware Low-Order Interaction Network (RLI) and a Network-Aware Attention Fusion Module (NAF). RLI explicitly preserves low-order signals while mitigating redundancy from residual connections, and NAF dynamically integrates explicit and implicit representations at each layer, enhancing information sharing and alleviating gradient imbalance. Together, these innovations balance low-order and high-order interactions, improving model expressiveness.
Extensive experiments on public datasets demonstrate that DLF achieves state-of-the-art performance in CTR prediction, addressing key limitations of existing models. The implementation is publicly available at \textcolor{blue}{\url{https://github.com/USTC-StarTeam/DLF}}.

\end{abstract}




\maketitle

\section{Introduction}

    \begin{figure}[t]
        \centering
        \setlength{\abovecaptionskip}{5pt}   
        \setlength{\belowcaptionskip}{0pt}   
        \includegraphics[width=0.45\textwidth]{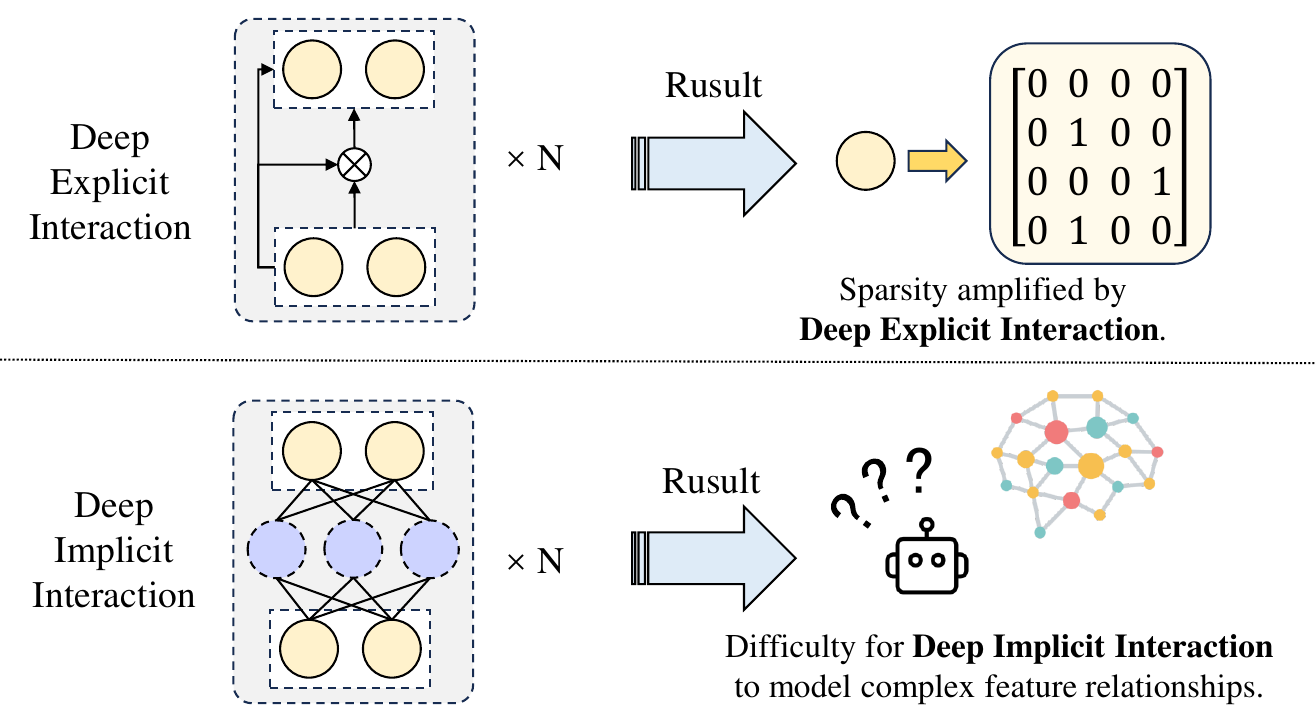}
        \caption{Drawbacks of deep explicit interaction and deep implicit interaction}
    \label{fig:motivation}
    \end{figure}

Recommender systems have become ubiquitous in digital platforms, helping users navigate through vast amounts of content by providing personalized suggestions based on their preferences and behaviors~\cite{tong2024mdap, zhang2024unified, yin2024learning, xie2024breaking, gu2025rapid}. Click-through rate (CTR) prediction is a fundamental task in online advertising and recommender systems~\cite{graepel2010web, he2014practical, LR, liu2023user, wang2025generative, wu2024survey}. It aims to estimate the probability that a user will click on a given item, such as an advertisement or a recommended product, based on the user's historical behavior~\cite{xu2024multi, wang2024denoising, han2024efficient}, item attributes~\cite{wang2025universal, wang2021decoupled, ye2025fuxi}, and contextual information~\cite{shen2024exploring}. Accurate CTR prediction is critical, as it directly affects the quality of recommendations, user engagement, and platform revenue. Typically, CTR prediction models take various features as input, encode them into dense representations, and then predict the likelihood of user clicks through learned feature interactions~\cite{DCN, DeepFM}.

A major challenge in CTR prediction lies in effectively modeling feature interactions, which are the relationships between different input features. These interactions can be categorized into explicit interactions and implicit interactions. Explicit interactions~\cite{FM, AFM} are those that are directly modeled or predefined, such as the interaction between two feature embeddings using mathematical operations like the inner product. Implicit interactions~\cite{FinalMLP}, on the other hand, are automatically learned by neural networks through non-linear transformations, without requiring explicit definitions. Both types of interactions are essential for accurately modeling user behavior in CTR tasks, as they provide complementary information about user preferences. 

Recent advances in feature interaction modeling have led to the development of two main paradigms: explicit interaction models and implicit interaction models. Explicit interaction models, such as FM~\cite{FM} and CIN~\cite{xDeepFM}, leverage inductive biases to efficiently capture low-order or high-order feature interactions. However, these methods often suffer from data sparsity issues\cite{zhang2025td3, zhang2025killingbirdsstoneunifying, han2023guesr, yin2024entropy, yin2024dataset}, which degrade their ability to learn robust high-level representations. On the other hand, implicit interaction models, such as multi-layer perceptrons (MLPs)~\cite{FinalMLP}, excel at capturing complex feature relationships through non-linear transformations. Yet, MLPs lack the inductive biases necessary for efficiently modeling simple operations like inner products~\cite{rendle2020neural}, which limits their ability to learn explicit patterns. This fundamental trade-off has motivated the development of two-stream models, such as DeepFM~\cite{DeepFM}, DCN~\cite{DeepFM}, and xDeepFM~\cite{xDeepFM}, which aim to integrate explicit and implicit interaction modeling into a unified framework.

Despite their success, two-stream models face several critical limitations. As shown in Figure~\ref{fig:motivation}, traditional two-stream architectures typically model explicit and implicit interactions independently, combining them only at the final layer. This separation leads to two major issues. First, the lack of interaction between the two streams prevents effective information sharing, which is particularly problematic as explicit streams struggle with data sparsity~\cite{deng2021deeplight, liu2020autofis}, while implicit streams fail to efficiently capture multiplicative feature interactions~\cite{rendle2020neural}. Second, the separation exacerbates gradient imbalance during backpropagation~\cite{hu2019dense}, where one stream may dominate the optimization process, hindering the overall convergence and expressiveness of the model. Furthermore, the effectiveness of residual connections in propagating low-order signals can be influenced by the unique characteristics of CTR data. Due to the inherent sparsity of CTR features and the potential limitations of feature embeddings, the representational quality of low-order signals may degrade as they are propagated through residual pathways. This leads to a compression of the vector space, where suboptimal embeddings occupy valuable representational capacity, thereby limiting the network's ability to model complex high-order interactions.

To address these limitations, we propose a novel framework, \textbf{\underline{D}ynamic \underline{L}ow-Order-Aware \underline{F}usion} \textbf{(DLF)}, which enables more effective and efficient explicit-implicit interaction modeling. The proposed framework introduces two key innovations: a \textbf{Residual-Aware Low-Order Interaction Network} \textbf{(RLI)} and a \textbf{Network-Aware Attention Fusion Module} \textbf{(NAF)}, which work together to overcome the challenges faced by existing two-stream models.

The RLI network is designed to explicitly preserve low-order signals while avoiding redundancy and interference caused by residual connections. Instead of relying solely on residual mechanisms to propagate low-order information, the RLI network introduces an additional pathway that explicitly captures low-order interactions between the initial feature embeddings and the current layer representations. This ensures that low-order features are preserved throughout the network, even as the depth of the interaction layers increases. To further address the potential conflict between residual signals and low-order interactions, a dynamic control mechanism is incorporated to adaptively balance their contributions. This design not only mitigates the sparsity issues faced by explicit streams but also enhances the overall expressiveness of the model by balancing low-order and high-order feature interactions.

In addition to the RLI network, we propose the NAF module to bridge the gap between explicit and implicit streams. Unlike traditional two-stream models that rely on late fusion, NAF facilitates layer-wise collaboration between the two streams by dynamically integrating their representations at each layer. This module employs a multi-level attention mechanism that captures both inter-block and intra-block feature interactions, ensuring that the fused representation retains the unique characteristics of each stream while modeling hierarchical relationships among feature interactions. By enabling effective information sharing between explicit and implicit streams, NAF alleviates gradient imbalance during optimization and improves the overall representation quality.
Our contributions are summarized as follows:
\begin{itemize}[leftmargin=*,align=left]
\item We propose a multi-level attention fusion mechanism that dynamically integrates explicit and implicit representations at each layer, enhancing information sharing and mitigating gradient imbalance during optimization.
\item We design a residual-aware low-order interaction network to alleviate sparsity issues in high-order interactions while preserving low-order features and avoiding noise amplification.
\item Extensive experiments on multiple public datasets demonstrate that our proposed method achieves state-of-the-art performance in CTR prediction tasks. Furthermore, the model exhibits strong adaptability, making it compatible with future advanced feature interaction modules.
\end{itemize}

\section{Related Work}

\subsection{Feature Interaction Modeling with Stacked Architectures}
Feature interaction modeling is a core challenge in CTR prediction, as it directly impacts a model's ability to capture user preferences and contextual relevance. Traditional methods, such as Logistic Regression (LR)~\cite{LR} and Factorization Machines (FM)~\cite{FM}, focus on low-order feature interactions but struggle to effectively capture high-order interactions, limiting their performance. With the advent of deep learning~\cite{guo2024scaling, shen2024optimizing, wang2025mf, wang2019mcne, wang2021hypersorec, xie2024bridging, yin2023apgl4sr, huang2024chemeval}, researchers have proposed various methods to enhance feature modeling through explicit and implicit interaction mechanisms. These methods can be broadly categorized based on how explicit and implicit interaction networks are integrated: stacked architectures and two-stream parallel architectures.

Stacked architectures adopt a sequential design, where explicit feature interactions are first captured by interaction networks, followed by deep neural networks (DNNs) to extract high-order implicit interactions. Representative explicit interaction mechanisms include operations based on the inner product, outer product, and Hadamard product, which are utilized in models like PNN~\cite{PNN}, OENN~\cite{OENN}, and NFM~\cite{NFM}. Other approaches, such as Cross Network (CN)~\cite{DCN} and its variants (e.g., CN-V2~\cite{DCNv2} and XCrossNet~\cite{DCNv3}), employ cross layers to model bounded-degree feature interactions. Additionally, attention-based methods, such as AutoInt~\cite{AutoInt}, AFM~\cite{AFM}, DIN~\cite{DIN}, and DIEN~\cite{DIEN}, leverage self-attention mechanisms to capture explicit interaction signals. 

More recent works have further explored novel designs for stacked architectures to enhance feature interaction modeling~\cite{AutoFis}. For instance, FiBiNet~\cite{FiBiNET} introduces a bilinear interaction layer combined with SENet~\cite{SENet} to adaptively recalibrate feature importance. Similarly, FAT-DeepFFM~\cite{FAT-DeepFFM} uses SENet to capture feature importance across fields. Another line of research focuses on leveraging tensor-based methods, such as TDM~\cite{TDM}, which utilize tensor decomposition techniques to efficiently model high-order feature interactions. Furthermore, methods like ONN~\cite{ONN} and DeepLight~\cite{DeepLight} emphasize the importance of interaction order by explicitly controlling the depth and complexity of feature combinations.
.

\subsection{Feature Interaction Modeling with Two-Stream Parallel Architectures}
In contrast to stacking architectures, two-stream models employ two parallel networks to simultaneously capture explicit and implicit feature interactions, which are then fused in the final layer. This design allows the two networks to focus on complementary aspects of feature interaction modeling. Early works, such as Wide \& Deep~\cite{WideDeep} and DeepFM~\cite{DeepFM}, adopt a dual-stream architecture, where the "wide" network captures low-order interactions (e.g., through LR~\cite{LR} or FM~\cite{FM}), while the "deep" network uses DNNs to model high-order implicit interactions. 

Subsequent studies have significantly advanced this paradigm. For example, xDeepFM~\cite{xDeepFM} incorporates Compressed Interaction Networks (CIN) to enhance explicit interaction modeling, while FinalMLP~\cite{FinalMLP} achieves competitive performance by stacking different numbers of MLP layers on each side. Other methods, such as HOFM~\cite{HOFM} and Deep \& Cross Network (DCN)~\cite{DCN}, focus on explicitly modeling high-order feature combinations, while maintaining the efficiency and scalability required for large-scale CTR prediction. Similarly, methods like FwFM~\cite{FwFM} and DIFM~\cite{DIFM} improve upon FM by introducing more flexible interaction mechanisms, such as field-aware and dynamic feature weighting.

A notable trend in recent works is the incorporation of attention mechanisms and graph-based designs into two-stream architectures. For instance, GraphFM~\cite{GraphFM} and NGCF~\cite{NGCF} leverage graph neural networks to capture higher-order feature dependencies, while maintaining a parallel structure to balance explicit and implicit interactions. Similarly, methods like AutoGroup~\cite{AutoGroup} and EDCN~\cite{EDCN} employ automated feature grouping techniques to improve interaction modeling. Another direction is the use of neural architecture search techniques, as seen in AutoCTR~\cite{AutoCTR}, to automatically design optimal two-stream architectures for different datasets.

Despite their advantages, two-stream architectures face challenges in effectively fusing explicit and implicit interactions. Improper fusion mechanisms may lead to overemphasis on one stream, resulting in suboptimal performance~\cite{FusionChallenge1, FusionChallenge2, FusionChallenge3}. Recent works, such as ESMM~\cite{ESMM}, DeepMCP~\cite{DeepMCP}, and MMoE~\cite{MMoE}, attempt to address this issue by introducing multi-task learning frameworks, which allow the model to jointly optimize multiple objectives while balancing the contributions of explicit and implicit networks.

\section{Preliminary}

\subsection{Problem Definition} 
Click-through rate prediction is a fundamental task in online recommendation systems, which aims to estimate the probability $\hat{y} = P(\text{click} \mid \mathbf{x})$ that a user will click on a given item. Here, $\mathbf{x} \in \mathbb{R}^n$ represents the concatenated input feature vector, combining both user and item features. The features typically include categorical attributes (e.g., user demographics, item categories) and numerical attributes (e.g., user activity statistics, item prices). CTR prediction tasks transform these high-dimensional sparse input features into low-dimensional dense embeddings $\mathbf{E}$ for representation. We define $y = 1$ to indicate a positive sample (click), and $y = 0$ to indicate a negative sample (no click). Based on the predicted probabilities $\hat{y}$, items are ranked to maximize user engagement.

\subsection{Feature Interaction in CTR Models}
In CTR prediction models, feature interaction is a critical mechanism for capturing relationships between features. Depending on how feature representations are combined, interactions can be categorized into two types: \textbf{interaction with the embedding layer} (low-order interactions) and \textbf{interaction within the current layer} (high-order interactions). These two approaches differ significantly in how they influence the growth of interaction order, as shown below.

\paragraph{Low-order interactions with the embedding layer.}
This approach computes interactions between the current layer's feature representations \(\mathbf{E}^{(l)}\) and the original embeddings \(\mathbf{E}^{(1)}\) from the embedding layer. The interaction is defined as:
\begin{equation}
\mathbf{z}_{ij}^{(l+1)} = \phi(\mathbf{e}_i^{(l)}, \mathbf{e}_j^{(1)}), \quad i, j \in \{1, 2, \ldots, N_l\},
\end{equation}
where \(\phi(\cdot, \cdot)\) is an interaction function (e.g., element-wise product, addition, or concatenation), \(N_l\) is the number of features in the \(l\)-th layer, and \(\mathbf{E}^{(1)} = \mathbf{E}\) represents the input embeddings. The resulting feature interaction order grows linearly with the layer index \(l\). For example, after \(l+1\) layers, the interaction order is proportional to:
\begin{equation}
\mathbf{x}^{(l+1)} = \psi(\mathbf{x}^{(l)}, \mathbf{x}^{(1)}),
\end{equation}
where \(\psi(\cdot)\) is an aggregation function. This linear growth ensures that low-order patterns are explicitly captured, but it limits the model's ability to represent complex, high-order interactions.

\paragraph{High-order interactions within the current layer.}
This approach computes interactions among the feature representations within the same layer, enabling the model to capture progressively higher-order relationships. The interaction is defined as:
\begin{equation}
\mathbf{z}_{ij}^{(l+1)} = \phi(\mathbf{e}_i^{(l)}, \mathbf{e}_j^{(l)}), \quad i, j \in \{1, 2, \ldots, N_l\}.
\end{equation}
In this case, the feature interaction order grows **exponentially** with the layer index \(l\). Specifically, after \(l+1\) layers, the interaction order is proportional to:
\begin{equation}
\mathbf{x}^{(2^{l+1})} = \psi(\mathbf{x}^{(2^l)}, \mathbf{x}^{(2^l)}),
\end{equation}
where \(\psi(\cdot)\) represents the interaction and aggregation function. This exponential growth allows the model to explore complex, high-order patterns, but it also increases the risk of overfitting and noise amplification.

\paragraph{The gap in interaction order.}
The key difference between these approaches lies in the growth of interaction order: low-order interactions grow linearly (\(l+1\)), while high-order interactions grow exponentially (\(2^{l+1}\)). This results in a substantial gap in their ability to capture complex feature relationships. Low-order interactions are effective for modeling simple, interpretable patterns but struggle with complex dependencies. High-order interactions, while powerful, may overemphasize abstract patterns, potentially overshadowing simpler relationships. However, fusing these two types of interactions is particularly challenging due to their differing growth rates and representational properties. The scale and distribution mismatch between the two types of interactions complicates the design of effective fusion mechanisms, often leading to optimization difficulties.

\section{Methodology}
    
    \begin{figure*}[ht]
        \centering
        \setlength{\abovecaptionskip}{5pt}   
        \setlength{\belowcaptionskip}{-6pt}   
        \includegraphics[width=0.9\textwidth]{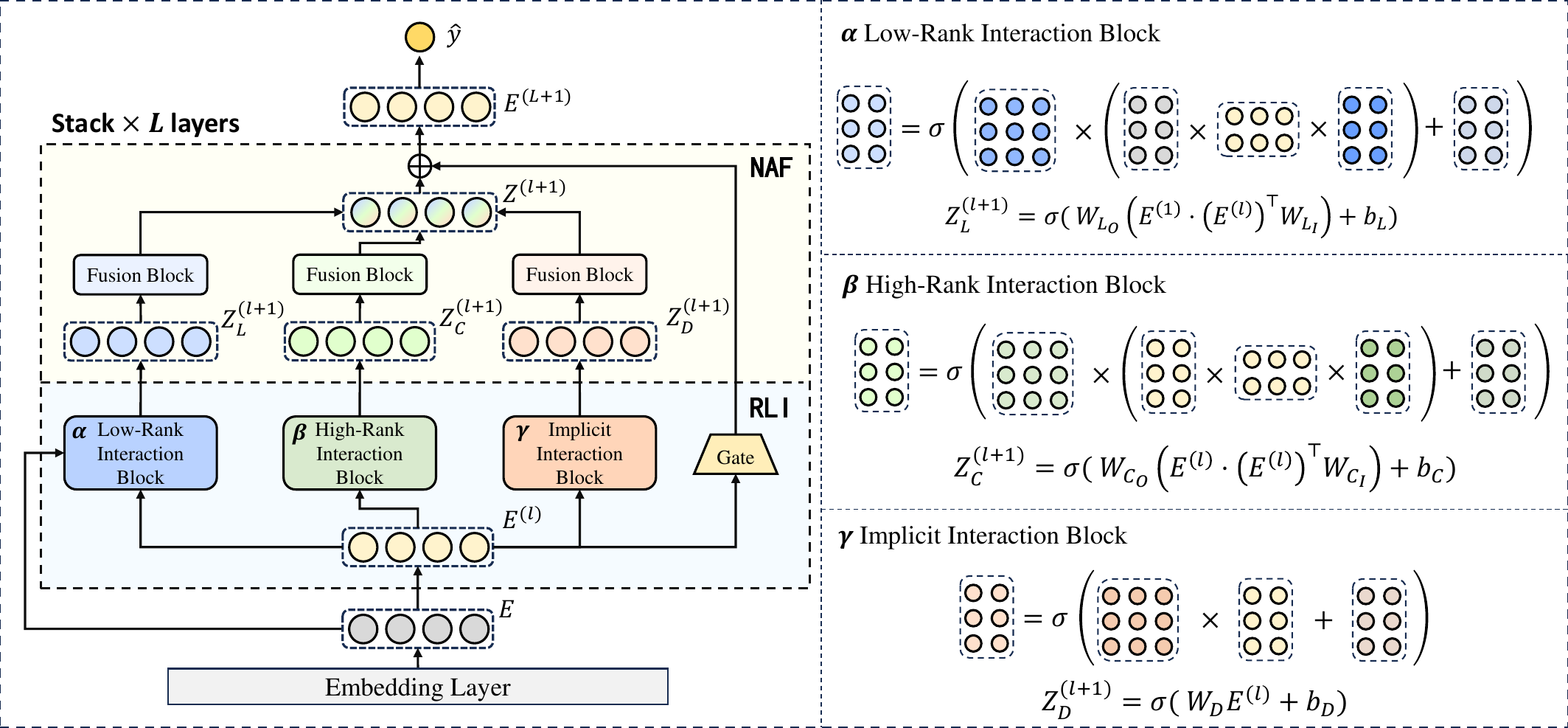}
        \caption{The overall framework of the proposed DLF model. The left part shows the complete workflow of DLF, with the core modules being the Residual-Aware Low-Order Interaction Network (RLI) and the Network-Aware Attention Fusion Module (NAF). The right part provides detailed visualizations of the three interaction blocks in RLI: the low-rank interaction block, the high-rank interaction block, and the implicit interaction block, highlighting their distinct computational mechanisms.}
    \label{fig:main}
    \end{figure*}

\subsection{Overview of the Proposed Method}

The proposed framework is designed to address the challenges of effectively modeling both low-order and high-order feature interactions in CTR prediction tasks while mitigating the suppression of low-order signals caused by deeper interaction layers. 
The workflow begins with an embedding layer that transforms the input features into dense vector representations, where categorical features are typically represented by trainable embedding vectors, while numerical features are normalized and directly embedded.
These embeddings \(\mathbf{E}\), serve as the starting point for the subsequent interaction modeling process.

The core of the framework consists of \(L\) stacked interaction layers, where each layer refines the feature representations by capturing diverse interaction patterns. At the \(l\)-th layer, the input representation \(\mathbf{E}^{(l)}\) is processed through the \textsl{Residual-Aware Low-Order Interaction Network}$~(\S~\ref{subsec:RLI})$, which includes three distinct pathways: (1) a low-rank interaction block that explicitly preserves low-order feature interactions, (2) a high-rank interaction block that captures higher-order feature interactions, and (3) an implicit interaction block that learns non-linear transformations in the feature space. The outputs of these pathways are denoted as \(\mathbf{Z}_L^{(l+1)}\), \(\mathbf{Z}_C^{(l+1)}\), and \(\mathbf{Z}_D^{(l+1)}\), are then fused using the \textsl{Network-Aware Attention Fusion Module}$~(\S~\ref{subsec:NAF})$. This fusion mechanism integrates both inter-block and intra-block feature interactions, producing the interaction-enhanced representation \(\mathbf{Z}^{(l+1)}\).

To ensure that essential low-order signals are retained throughout the network, a gated residual mechanism combines the interaction-enhanced representation \(\mathbf{Z}^{(l+1)}\) with the input of the current layer \(\mathbf{E}^{(l)}\). This residual connection not only stabilizes the learning process but also prevents the degradation of low-order feature information as the network depth increases. The resulting representation, \(\mathbf{E}^{(l+1)}\), serves as the input for the next layer.

As described in $~(\S~\ref{subsec:predict})$, after \(L\) layers of interaction modeling, the final representation \(\mathbf{E}^{(L+1)}\) is passed through a fully connected prediction network. A sigmoid activation function is applied to compress the output into a single scalar value, \(\hat{y}\), which represents the predicted click-through probability. The entire framework is trained by minimizing the binary cross-entropy loss, ensuring that the model learns to accurately predict user interactions based on the input features.

\subsection{Residual-Aware Low-Order Interaction Network}
\label{subsec:RLI}

In deep CTR prediction models, feature representations from the \(l\)-th layer, denoted as \(\mathbf{E}^{(l)}\), are typically processed through both explicit and implicit interaction layers to capture higher-order feature interactions. These layers generate an interaction-enhanced representation, which is then combined with a residual connection to form the updated feature representation for the next layer. While the residual connection ensures that low-order signals from the previous layer are preserved, it does not explicitly address the suppression of these signals caused by noise amplification in higher-order interactions. As interaction depth increases, the sparsity of the feature space grows exponentially, leading to the loss of low-order signals and degradation of model performance.

For this, we propose the \textbf{Residual-Aware Low-Order Interaction Network}, which introduces multiple pathways to effectively capture low-order, high-order, and implicit feature interactions. These pathways include the low-rank interaction block, high-rank interaction block, and implicit interaction block, each designed to focus on different aspects of feature interaction modeling.

The first pathway, referred to as the low-rank interaction block, interacts with the initial feature embeddings \(\mathbf{E}^{(1)}\) and the current layer representations \(\mathbf{E}^{(l)}\) to explicitly capture low-order interactions. The output of this pathway is computed as:
\begin{equation}
\mathbf{Z}_L^{(l+1)} = \sigma(\mathbf{W}_{L_O} (\mathbf{E}^{(1)} \cdot (\mathbf{E}^{(l)})^\top \mathbf{W}_{L_I} ) + \mathbf{b}_L),
\end{equation}
where \(\mathbf{W}_{L_I} \in \mathbb{R}^{R \times d}\) is the transformation matrix applied to the current layer representations, \(\mathbf{W}_{L_O} \in \mathbb{R}^{d \times d}\) is the weight matrix applied to the low-order interaction output, \(\mathbf{b}_L \in \mathbb{R}^d\) is the bias term, and \(\sigma(\cdot)\) is the ReLU activation function. By interacting \(\mathbf{E}^{(1)}\) with \(\mathbf{E}^{(l)}\), this pathway ensures that low-order signals are preserved throughout the interaction process, even as network depth increases.

Simultaneously, the current layer representations \(\mathbf{E}^{(l)}\) are processed through a second pathway to model higher-order interactions. This pathway, referred to as the high-rank interaction block, explicitly captures higher-order interactions using a CrossNet-inspired architecture. The output of this pathway is:
\begin{equation}
\mathbf{Z}_C^{(l+1)} = \sigma(\mathbf{W}_{C_O} (\mathbf{E}^{(l)} \cdot (\mathbf{E}^{(l)})^\top \mathbf{W}_{C_I} ) + \mathbf{b}_C),
\end{equation}
where \(\mathbf{W}_{C_I} \in \mathbb{R}^{R \times d}\) is the transformation matrix applied to the higher-order interaction input, \(\mathbf{W}_{C_O} \in \mathbb{R}^{d \times d}\) is the weight matrix applied to the higher-order interaction output, and \(\mathbf{b}_C \in \mathbb{R}^d\) is the bias term. By repeatedly interacting \(\mathbf{E}^{(l)}\) with itself, this pathway captures progressively more complex feature interactions as the network deepens.

In addition to the explicit interaction pathways, implicit interactions are modeled through a dense transformation layer, referred to as the implicit interaction block. This block is implemented as a multi-layer perceptron (MLP) with ReLU activations. The output of this pathway is computed as:
\begin{equation}
\mathbf{Z}_D^{(l+1)} = \sigma(\mathbf{W}_D \mathbf{E}^{(l)} + \mathbf{b}_D),
\end{equation}
where \(\mathbf{W}_D \in \mathbb{R}^{d \times d}\) and \(\mathbf{b}_D \in \mathbb{R}^d\) are the weight matrix and bias term, respectively. Unlike the explicit interaction pathways, this transformation layer focuses on learning implicit patterns in the feature space through non-linear transformations.

The outputs of these pathways, \(\mathbf{Z}_L^{(l+1)}\), \(\mathbf{Z}_C^{(l+1)}\), and \(\mathbf{Z}_D^{(l+1)}\), are then fused to form the overall interaction-enhanced representation:
\begin{equation}
\mathbf{Z}^{(l+1)} = \psi(\mathbf{Z}_L^{(l+1)}, \mathbf{Z}_C^{(l+1)}, \mathbf{Z}_D^{(l+1)}).
\end{equation}

However, the interaction of low-order signals through the LOPL and the implicit retention of these signals via the residual connection may lead to redundancy or conflict. To address this, we introduce a \textbf{gated residual mechanism} that adaptively balances the contributions of the residual connection and the interaction-enhanced representation. Specifically, the updated feature representation is computed as:

\begin{equation}
\mathbf{E}^{(l+1)} = \mathbf{Z}^{(l+1)} + \max(\epsilon, \mathbf{W}_\text{Gate} \mathbf{E}^{(l)}) \odot \mathbf{E}^{(l)},
\end{equation}
where \(\mathbf{W}_\text{Gate} \in \mathbb{R}^{d \times d}\) is a learnable gating matrix, \(\epsilon\) is a small positive constant to ensure numerical stability, and \(\odot\) denotes element-wise multiplication. The gating mechanism dynamically adjusts the residual connection's contribution based on the importance of the current layer's representations, ensuring that the residual connection complements rather than conflicts with the low-order preservation pathway.

By introducing the RLI and the gated residual mechanism, the proposed framework effectively addresses the suppression of low-order signals in high-order interaction modeling. This design ensures that both low-order and high-order patterns are captured in a balanced and hierarchical manner, enhancing the model's ability to learn complex feature relationships for CTR prediction.

\subsection{Network-Aware Attention Fusion Module}
\label{subsec:NAF}

To effectively integrate the outputs of the proposed Residual-Aware Low-Order Interaction Network, we introduce a \textbf{Network-Aware Attention Fusion Module}. It is designed to capture both inter-block and intra-block feature interactions, ensuring that the fused representation retains the unique characteristics of each block while modeling the hierarchical relationships between them. The following operations are applied to the block outputs at each layer, where the representations \(\mathbf{Z}_L, \mathbf{Z}_C, \mathbf{Z}_D\) are assumed to represent the outputs of any given layer and the process is repeated for all layers.

\subsubsection{Block-Wise Attention Mechanism}

To model the interactions between different blocks, we first compute the query, key, and value matrices for all three blocks (\(\mathbf{Z}_L, \mathbf{Z}_C, \mathbf{Z}_D\)) in a block-wise manner. Specifically, we compute:
\begin{equation}
    (\mathbf{Q}_M, \mathbf{K}_M, \mathbf{V}_M) = \text{Split}(f_{\text{att}}(\mathbf{Z}_M)), M \in \{L, C, D\}
\end{equation}
where \(f_{\text{att}}(\cdot)\) represents the operation that constructs the Query, Key, and Value from the input in attention blocks.

For inter-block cross-attention, the query matrix of one block (\(M\)) is used to compute attention scores with the key matrices of the other two blocks (\(X\) and \(Y\)). The cross-attention outputs are computed as:
\begin{equation}
\mathbf{A}_{MX}^\text{C} = \text{softmax}\left(\frac{\mathbf{Q}_M \mathbf{K}_X^\top}{\sqrt{d}}\right), \quad \mathbf{A}_{MY}^\text{C} = \text{softmax}\left(\frac{\mathbf{Q}_M \mathbf{K}_Y^\top}{\sqrt{d}}\right),
\end{equation}
\begin{equation}
\text{Att}_M^\text{C} = \sigma\left(\mathbf{W}_M^\text{C} [\mathbf{A}_{MX}^\text{C} \mathbf{V}_X; \mathbf{A}_{MY}^\text{C} \mathbf{V}_Y] + \mathbf{b}_M^\text{C}\right),
\end{equation}
where \(M \in \{L, C, D\}\) represents the current block, and \(X, Y\) are the other two blocks (\(X, Y \in \{L, C, D\} \setminus \{M\}\)). \([\cdot; \cdot]\) denotes concatenation, \(\mathbf{W}_M^\text{C} \in \mathbb{R}^{d \times 2d}\), and \(\mathbf{b}_M^\text{C} \in \mathbb{R}^d\) are learnable parameters.

In addition to inter-block interactions, we also model intra-block feature interactions using self-attention. For each block \(M\), the self-attention is computed as:
\begin{equation}
\mathbf{A}_{M}^\text{S} = \text{softmax}\left(\frac{\mathbf{Q}_M \mathbf{K}_M^\top}{\sqrt{d}}\right), \quad \text{Att}_M^\text{S} = \mathbf{A}_M^\text{S} \mathbf{V}_M,
\end{equation}
where \(\mathbf{Q}_M, \mathbf{K}_M, \mathbf{V}_M\) are the query, key, and value matrices corresponding to block \(M\). This allows the model to capture fine-grained feature interactions within each block.

\subsubsection{Fusion and Output Generation}

To produce the fused attention representation for each block, we aggregate the cross-attention and self-attention outputs. Specifically, the attention representation for each block is computed as:
\begin{equation}
\text{Att} = \text{ReLU}\left(\mathbf{W}_F \left[\text{Att}_L^\text{C} + \text{Att}_L^\text{S}; \text{Att}_C^\text{C} + \text{Att}_C^\text{S}; \text{Att}_D^\text{C} + \text{Att}_D^\text{S}\right] + \mathbf{b}_F\right),
\end{equation}
where \(\mathbf{W}_F \in \mathbb{R}^{d \times 3d}\) and \(\mathbf{b}_F \in \mathbb{R}^d\) are learnable parameters, and \([\cdot; \cdot; \cdot]\) denotes concatenation across blocks.

Finally, the fused attention representations are concatenated with the original block outputs to form the input for the next layer:
\begin{equation}
\mathbf{Z} = \left[\mathbf{Z}_L; \mathbf{Z}_C; \mathbf{Z}_D; \text{Att}\right].
\end{equation}

This design ensures that both intra-block and inter-block interactions are effectively captured, while preserving the original block-specific outputs for hierarchical feature learning.

\subsection{Prediction and Optimization}
\label{subsec:predict}
After obtaining the final layer representation \(\mathbf{E}^{(L+1)}\), we utilize a fully connected layer followed by a sigmoid activation function to compress the representation into a single scalar value \(\hat{y}\), which represents the predicted probability of the positive class. The prediction process is formulated as follows:

\begin{equation}
\hat{y} = \sigma(\mathbf{W}_\text{out}^\top \mathbf{E}^{(L+1)} + b_\text{out}),
\end{equation}
where \(\mathbf{W}_\text{out} \in \mathbb{R}^d\) is the learnable weight vector, \(b_\text{out} \in \mathbb{R}\) is the bias term, and \(\sigma(\cdot)\) denotes the sigmoid activation function.

To train the model, we optimize the binary cross-entropy (BCE) loss between the predicted probability \(\hat{y}\) and the ground truth label \(y \in \{0, 1\}\). The BCE loss is defined as:

\begin{equation}
\mathcal{L}_\text{BCE} = - \frac{1}{N} \sum_{i=1}^N \left[ y_i \log(\hat{y}_i) + (1 - y_i) \log(1 - \hat{y}_i) \right],
\end{equation}
where \(N\) is the number of training samples, \(y_i\) is the ground truth label for the \(i\)-th sample, and \(\hat{y}_i\) is the predicted probability for the \(i\)-th sample.

\section{EXPERIMENTS}

    \begin{table}[t]
    \caption{Dataset statistics}
    \centering
    \vspace{-0.2cm}
    \setlength{\tabcolsep}{1mm}
    \begin{tabular}{c|c|c|c}
        \toprule
        \textbf{Dataset} & \textbf{\#Interactions} & \textbf{\#Feature Fields} & \textbf{\#Features} \\
        \midrule
        Criteo & 45,840,617 & 39 & 1,086,810 \\
        Avazu & 40,428,968 & 22 & 2,018,012 \\
        Movielens & 2,006,859 & 3 & 90,242 \\
        Frappe & 288,609 & 10 & 5,339 \\
        \bottomrule
    \end{tabular}
    \label{tab:dataset}
\end{table}
    \subsection{Experiment Setup}
        \subsubsection{Datasets.}
            To evaluate the performance of our proposed models, we utilized four different datasets from the real world. 

            \begin{itemize}[leftmargin=*,align=left]
                \item \textbf{Criteo}\footnote{https://www.kaggle.com/c/criteo-display-ad-challenge/}: An industry-standard large-scale dataset for ad click-through rate (CTR) prediction, containing 45 million samples with 13 continuous and 26 categorical features.
            
                \item \textbf{Avazu}\footnote{https://www.kaggle.com/c/avazu-ctr-prediction/}: A benchmark dataset for CTR prediction, consisting of 40 million samples and 24 features collected over 11 days.
            
                \item \textbf{MovieLens}\footnote{https://grouplens.org/datasets/movielens}: A dataset for personalized tag recommendation, containing 2 million interactions represented as \textit{(user\_id, item\_id, tag\_id)} with 90,242 unique feature values.
            
                \item \textbf{Frappe}\footnote{https://www.baltrunas.info/context-aware}: A context-aware app usage dataset with 96,203 interactions across 4,082 apps, represented by 10 feature fields including \textit{user\_id}, \textit{item\_id}, and contextual features.
            \end{itemize}

            We followed the preprocessing steps outlined in AFN \cite{AFN} to split
            all the datasets into training, validation, and test sets in a 7:2:1 ratio, following the time order.
            Table \ref{tab:dataset} provides detailed statistics for these two public datasets. 
            
    \subsubsection{Baseline Models}
    To demonstrate the effectiveness of the proposed model, we select some representative CTR models for comparison.
    We also compare our proposed model with some hashing and quantization-based methods to validate the superiority of our model. 
    The details are listed as follows:

\textbf{Explicit Feature Interaction Models}
Explicit feature interaction models focus on directly modeling feature interactions through predefined mechanisms, such as factorization or attention, to capture explicit relationships without relying on implicit representations. Representative models include:
\begin{itemize}[leftmargin=*,align=left]
    \item \textbf{FM} \cite{FM}: Factorization Machine models pairwise interactions between features efficiently, making it suitable for sparse data.
    \item \textbf{AFM} \cite{AFM}: Attentional Factorization Machine extends FM by introducing an attention mechanism to quantify the importance of feature interactions.
    \item \textbf{SAM} \cite{SAM}: Self-Attentive Model leverages self-attention to explicitly capture complex feature interactions with improved flexibility and effectiveness.
    \item \textbf{AutoInt} \cite{AutoInt}: Automatic Feature Interaction Learning via Self-Attention employs multi-head self-attention to automatically learn explicit feature interactions without manual engineering.
\end{itemize}

\textbf{Dual-Stream Explicit and Implicit Feature Interaction Fusion Models}
Dual-stream models integrate explicit and implicit feature interactions into a unified framework, enabling the simultaneous modeling of low-order explicit interactions and high-order implicit relationships. These models are designed to balance interpretability and predictive accuracy. Representative models include:
\begin{table*}[t]
\setlength{\abovecaptionskip}{0.05cm}
\setlength{\belowcaptionskip}{0.2cm}
\caption{Overall performance comparison between the baselines and DLF across four datasets.
Bold values indicate a statistically significant level $p$-value<0.05 comparing our DLF with the base model's performance in terms of AUC and LogLoss.}
\setlength{\tabcolsep}{4mm}{
\resizebox{0.9\textwidth}{!}{
\begin{tabular}{c|c|c|c|c|c|c|c|c|c}
\toprule
    \multirow{2}{1cm}[-1pt]{\centering\textbf{Model Class}} & 
    \multirow{2}{1cm}[-1pt]{\centering\textbf{Model}} & 
    \multicolumn{2}{c|}{\multirow{-1}{*}{\textbf{Avazu}}} &
    \multicolumn{2}{c|}{\multirow{-1}{*}{\textbf{Criteo}}} &
    \multicolumn{2}{c|}{\multirow{-1}{*}{\textbf{Movielens}}} &
    \multicolumn{2}{c}{\multirow{-1}{*}{\textbf{Frappe}}} \\ \cline{3-10}
    & & 
    \multicolumn{1}{c|}{\textbf{AUC}} &
    \multicolumn{1}{c|}{\textbf{Logloss}} &
    \multicolumn{1}{c|}{\textbf{AUC}} &
    \multicolumn{1}{c|}{\textbf{Logloss}} &
    \multicolumn{1}{c|}{\textbf{AUC}} &
    \multicolumn{1}{c|}{\textbf{Logloss}} &
    \multicolumn{1}{c|}{\textbf{AUC}} &
    \multicolumn{1}{c}{\textbf{Logloss}} \\   \midrule\midrule

    \multirow{4}{1.5cm}{\centering \textbf{Explicit Interaction}} & \textbf{FM} & 0.7509  & 0.3769  & 0.7931  & 0.4587  & 0.9424  & 0.2737  & 0.9609  & 0.2297   \\ 
    &\textbf{AFM} & 0.7526  & 0.3719  & 0.7966  & 0.4540  & 0.9451  & 0.2658  & 0.9602  & 0.2218   \\ 
    &\textbf{SAM} & 0.7527  & 0.3722  & 0.7990  & 0.4523  & 0.9491  & 0.2617  & 0.9606  & 0.2216   \\ 
    &\textbf{AutoInt} & 0.7539  & 0.3740  & 0.8084  & 0.4435  & 0.9430  & 0.2705  & 0.9680  & 0.2231   \\ \midrule
    \multirow{8}{1cm}{\centering \textbf{Dual Stream Fusion}} & \textbf{DCN} & 0.7537  & 0.3745  & 0.8085  & 0.4428  & 0.9448  & 0.2708  & 0.9687  & 0.2161   \\ 
    &\textbf{DCNv2} & 0.7565  & 0.3722  & 0.8085  & 0.4431  & 0.9423  & 0.2774 & 0.9717  & 0.2137    \\ 
    &\textbf{DeepFM} & 0.7540  & 0.3742  & 0.8087  & 0.4462  & 0.9428  & 0.2800  & 0.9688  & 0.2139   \\ 
    &\textbf{xDeepFM} & 0.7532  & 0.3786  & 0.8088  & 0.4436  & 0.9418  & 0.2792  & 0.9691  & 0.2287   \\ 
    &\textbf{FinalNet} & 0.7577  & 0.3711  & 0.8088  & 0.4430  & 0.9495  & 0.2608  & 0.9788  & \underline{0.1810}   \\ 
    &\textbf{FinalMLP} & 0.7533  & 0.3741  & 0.8088  & 0.4429  & 0.9438  & 0.2741  & 0.9607  & 0.2399   \\ 
    &\textbf{GDCN} & 0.7529  & 0.3750  & 0.8095  & \underline{0.4414}  & 0.9416  & 0.2785  & 0.9615  & 0.2307   \\ 
    &\textbf{PNN} & 0.7543  & 0.3741  & \underline{0.8096}  & 0.4416  & 0.9472  & 0.2745  & 0.9743  & 0.2116   \\ \midrule
    \multirow{3}{1cm}{\centering \textbf{Layer Wise Fusion}}
    &\textbf{EDCN} & 0.7578  & \underline{0.3709}  & 0.8092  & 0.4418  & 0.9472  & 0.2357  & 0.9649  & 0.2357   \\ 
    & \textbf{Wukong} & \underline{0.7580}  & 0.3715  & 0.8091  & 0.4424  & \underline{0.9502}  & \underline{0.2285}  & \underline{0.9794}  & 0.1867   \\ 
    &\textbf{Ours} & \textbf{0.7601}  & \textbf{0.3706}  & \textbf{0.8106}  & \textbf{0.4412}  & \textbf{0.9515}  & \textbf{0.2223}  & \textbf{0.9812}  & \textbf{0.1679}   \\ \bottomrule

\end{tabular}
}}
\label{tab:MainTable}
\end{table*} 
\begin{itemize}[leftmargin=*,align=left]
    \item \textbf{DeepFM} \cite{DeepFM}: Combines FM for explicit interactions and DNN for implicit interactions, improving recommendation accuracy.
    \item \textbf{PNN} \cite{PNN}: Product-based Neural Network explicitly models feature interactions using product operations on feature embeddings while leveraging neural networks for implicit interactions.
    \item \textbf{DCN} \cite{DCN}: Captures explicit interactions via cross layers and implicit interactions via deep layers, balancing interpretability and expressiveness.
    \item \textbf{DCNv2} \cite{DCNv2}: Extends DCN by stacking multiple cross and deep layers for enhanced interaction modeling.
    \item \textbf{AFN} \cite{AFN}: Dynamically learns feature interaction importance using adaptive factorization.
    \item \textbf{GDCN} \cite{GDCN}: Uses gating mechanisms to model complex user-item relationships, enhancing collaborative filtering signals.
    \item \textbf{FinalMLP} \cite{FinalMLP}: Combines two MLP networks with gating and aggregation layers for effective stream-level interaction integration to enhance feature input.
    \item \textbf{FinalNet} \cite{FinalNet}: Uses factorized interaction layers to efficiently learn high-order interactions, forming a unified backbone for CTR models.
    \item \textbf{xDeepFM} \cite{xDeepFM}: Introduces a Compressed Interaction Network (CIN) to model high-order interactions efficiently and interpretably, while leveraging DNNs for implicit interactions.
\end{itemize}

\textbf{Layer-Wise Explicit and Implicit Feature Interaction Fusion Models}
Layer-wise explicit and implicit feature interaction fusion models focus on progressively integrating explicit and implicit interactions across layers, enhancing cross-layer feature learning and improving information sharing. Representative models include:
\begin{itemize}[leftmargin=*,align=left]
    \item \textbf{EDCN} \cite{EDCN}: Enhanced Deep \& Cross Network addresses the insufficient and excessive sharing issues in parallel deep CTR models (e.g., DCN). It introduces a bridge module and a regulation module to enhance information sharing between explicit and implicit feature interactions, improving layer-wise feature learning and model expressiveness.
    \item \textbf{Wukong} \cite{Wukong}: A novel layer-wise feature interaction model that dynamically integrates explicit and implicit feature representations across layers, leveraging attention mechanisms and gating modules to achieve fine-grained feature interaction learning.
\end{itemize}

    \subsubsection{Evaluation Metrics.}

            We evaluate the algorithms using two popular metrics: AUC \cite{AUC} and Logloss \cite{LogicR}. The AUC (Area Under the ROC Curve) metric measures the ability of the model to rank positive items (i.e., samples with label 1) higher than negative ones. A higher AUC indicates better recommendation performance. Logloss measures the distance between the predicted probabilities and the ground-truth labels.

        \subsubsection{Parameter Settings.}
            We implemented all models utilizing FuxiCTR, an open-source CTR prediction library \footnote{https://fuxictr.github.io/}. 
            To ensure a fair comparison, we standardized the embedding dimension across all models to 64 and set the batch size to 10,000. 
            The learning rate was tuned from the set {1e-1, 1e-2, 1e-3, 1e-4}, while $L_2$ regularization was adjusted within the range {0, 1e-1, 1e-2, 1e-3, 1e-4, 1e-5}, and the dropout ratio varied from 0 to 0.9. 
            All baseline models were trained using the Adam \cite{adam} optimizer. 
            To ensure the robustness and reliability of our findings, each experiment was conducted five times, with the average performance being reported. 

    \subsection{Performance Comparison}

        We conducted extensive experiments to evaluate the performance of various models across multiple datasets, focusing on their ability to model feature interactions effectively. The models are categorized into three groups based on their interaction modeling strategies: (i) explicit interaction models, including FM, AFM, SAM, and AutoInt; (ii) dual-stream models that combine explicit and implicit feature interactions, such as DCN, DCNv2, DeepFM, xDeepFM, FinalNet, FinalMLP, GDCN, and PNN; and (iii) layer-wise fusion models, including Wukong, EDCN, and our proposed method. Below, we summarize the key findings from our experiments:

        \textbf{(1) Models leveraging both explicit and implicit feature interactions outperform those using only explicit interactions.}
        The results consistently demonstrate that dual-stream models and layer-wise fusion models achieve better performance compared to models that rely solely on explicit feature interactions. For instance, while FM and AFM provide a solid baseline by explicitly modeling second-order interactions, they fail to capture higher-order or implicit interactions effectively, leading to a performance gap when compared to models like DeepFM or DCNv2. This highlights the importance of complementing explicit interaction modeling with implicit feature extraction, as it enables richer representations of complex data patterns.

        \textbf{(2) Layer-wise fusion of explicit and implicit interactions yields superior results compared to traditional dual-stream models.}
        Our experiments reveal that models employing layer-wise fusion, such as Wukong, EDCN, and our proposed method, consistently outperform traditional dual-stream models that integrate explicit and implicit interactions only at the final stage. For example, even Wukong, which adopts a relatively simple structure, achieves competitive performance compared to more complex dual-stream models like xDeepFM and DCNv2. This suggests that progressively fusing explicit and implicit interactions at each layer enables a more effective hierarchical representation of features, leading to improved predictive accuracy in CTR prediction. Notably, our method achieves state-of-the-art performance, further validating the effectiveness of layer-wise fusion.

        \textbf{(3) The proposed method achieves significant improvements across all datasets, particularly on those with sparse features.}
        Our model consistently outperforms all baselines on every dataset, demonstrating its robustness and generalization capability. The improvement is particularly pronounced on datasets with smaller sample sizes or sparser features, such as Frappe, where traditional models often struggle due to the lack of sufficient data to learn meaningful representations. This result underscores the strength of our approach in modeling sparse and high-dimensional feature spaces, which is achieved through its ability to effectively integrate explicit and implicit interactions in a layer-wise manner. Furthermore, the performance gap between our model and others widens as the sparsity of the dataset increases, highlighting its adaptability to challenging scenarios.

    
        \subsection{\textbf{Ablation Study}}
            \begin{table}[t]
\setlength{\abovecaptionskip}{0.05cm}
\setlength{\belowcaptionskip}{0.2cm}
\caption{Ablation study of DLF}
\setlength{\tabcolsep}{3mm}{
\resizebox{0.45\textwidth}{!}{
\begin{tabular}{c|c|c|c|c}
\toprule
 &
    \multicolumn{2}{c|}{\multirow{-1}{*}{\textbf{Criteo}}} &
    \multicolumn{2}{c}{\multirow{-1}{*}{\textbf{Avazu}}} \\ \midrule
    \multicolumn{1}{c|}{\textbf{Variants}} &
    \multicolumn{1}{c|}{\textbf{AUC}} &
    \multicolumn{1}{c|}{\textbf{Logloss}} &
    \multicolumn{1}{c|}{\textbf{AUC}} &
    \multicolumn{1}{c}{\textbf{Logloss}} \\   \midrule\midrule
    \textbf{$\text{DLF}$} & \textbf{0.8106} & \textbf{0.4412} & \textbf{0.7601} & \textbf{0.3706} \\
    \textbf{w/o gate} & 0.8095 & 0.4422 & 0.7591 & 0.3719 \\
    \textbf{w/o RLI} & 0.8091 & 0.4428 & 0.7590 & 0.3721 \\
    \textbf{w/o NAF} & 0.8083 & 0.4438 & 0.7575 & 0.3726 \\
    \textbf{w/o RLI + NAF} & 0.8084 & 0.4437 & 0.7563 & 0.3731 \\
\bottomrule
\end{tabular}
}}
\label{tab:Ablation}
\end{table}
To evaluate the contribution of each component in our proposed model, we conduct a series of ablation experiments by selectively removing key modules and comparing their performance. Specifically, we analyze five configurations: (1) the full model, (2) the model without residual gating, (3) the model without the Residual-Aware Low-Order Interaction (RLI) network, retaining only conventional explicit and implicit interaction modules, (4) the model without the Network-Aware Attention Fusion (NAF) block, where fusion is only performed in the final layer as in traditional two-stream models, and (5) the model with both RLI and NAF removed. The results are summarized in Table~\ref{tab:Ablation}.
            
From the results, we observe that removing the residual gating mechanism leads to a noticeable performance drop, highlighting its importance in filtering out noisy low-order information. Interestingly, when the RLI module is further removed, the performance degradation is slightly mitigated, suggesting that in the absence of residual gating, the low-order information introduced by RLI may conflict with residual connections. By removing RLI entirely, the model avoids this conflict but loses the ability to effectively model low-order interactions. This observation underscores the critical role of residual gating in retaining meaningful low-order information and integrating it seamlessly into the model.
            
On the other hand, the removal of NAF results in a significant performance drop, which is only marginally better than removing both RLI and NAF. This result highlights the importance of intra-layer fusion in mitigating sparsity issues associated with high-order interactions. Without this part, the model suffers from suboptimal feature fusion, leading to degraded performance.
            
In summary, the ablation study demonstrates the complementary roles of RLI and NAF. RLI captures and refines low-order interactions, while NAF ensures effective information flow across different levels of interactions. Together, these components enable our model to achieve superior performance compared to traditional feature interaction models.

        \subsection{Analysis of Fusion Methods}
            \begin{figure}[t]
                \centering
                \includegraphics[width=0.5\textwidth]{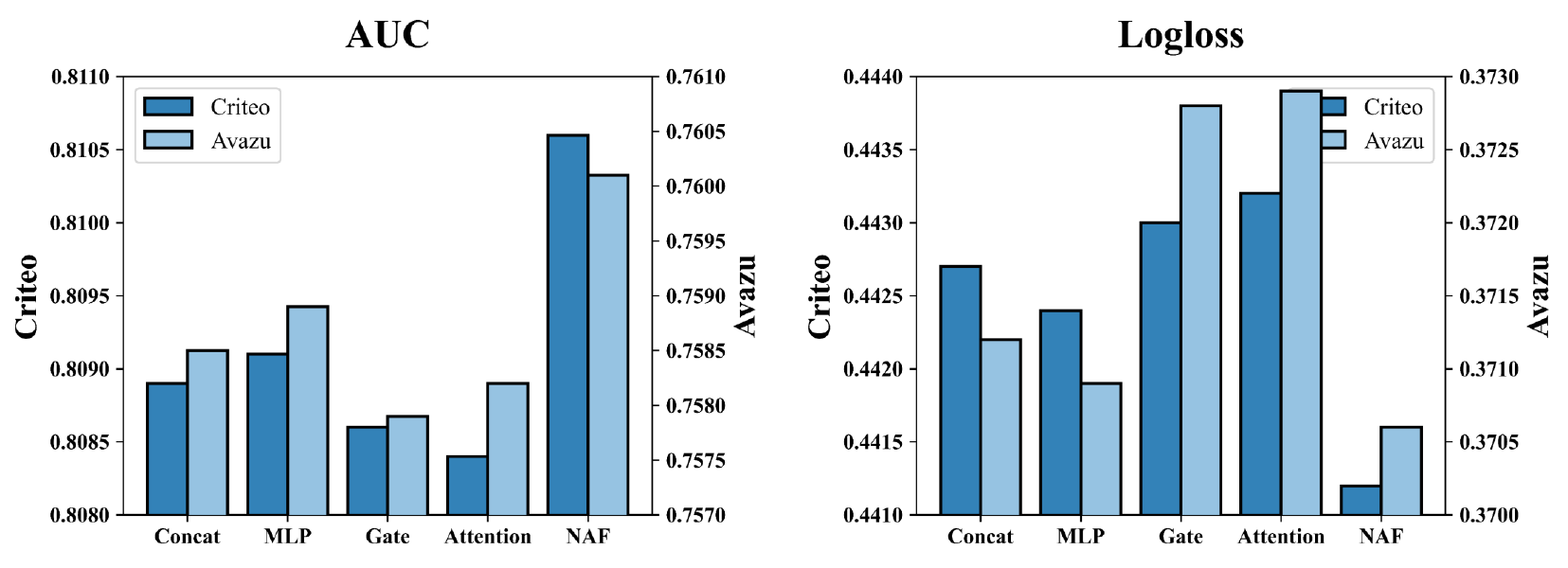}
                \vspace{-10pt}
                \caption{Interaction fusion methods analysis}
            \label{fig:zhu_chart}
            \end{figure}

        To explore the effectiveness of different fusion strategies, we conducted experiments comparing five methods: direct concatenation, MLP-based fusion, gate-pooling, attention-pooling, and our proposed approach. The results, as shown in Figure~\ref{fig:zhu_chart}, reveal significant performance differences among these methods.
        
        Direct concatenation, the simplest approach, combines embeddings without further interaction. While computationally efficient, it fails to capture meaningful relationships, resulting in the lowest performance. MLP-based fusion introduces non-linear transformations for deeper feature interactions, achieving better results. Gate-pooling applies learnable gating weights to embeddings but is limited by its simplistic mechanism, leading to inferior performance compared to MLP-based fusion. Attention-pooling assigns data-dependent importance scores via cross-attention, enabling fine-grained interactions, but its focus on cross-module interactions neglects internal consistency, limiting its performance gains.
        
        In contrast, our proposed method achieves the best performance by combining cross-module attention with intra-module self-attention. This dual mechanism facilitates effective information exchange while preserving and refining module-specific information. As shown in Figure~\ref{fig:line_chart}, our method significantly outperforms all others, highlighting the importance of balancing cross-module interaction and intra-module consistency. These findings emphasize the necessity of advanced fusion mechanisms to fully utilize the hierarchical and complementary nature of embeddings.

        \subsection{\textbf{Analysis of Hyper-Parameter}}              
\begin{figure}[t!]
                \centering
            \includegraphics[width=0.5\textwidth]{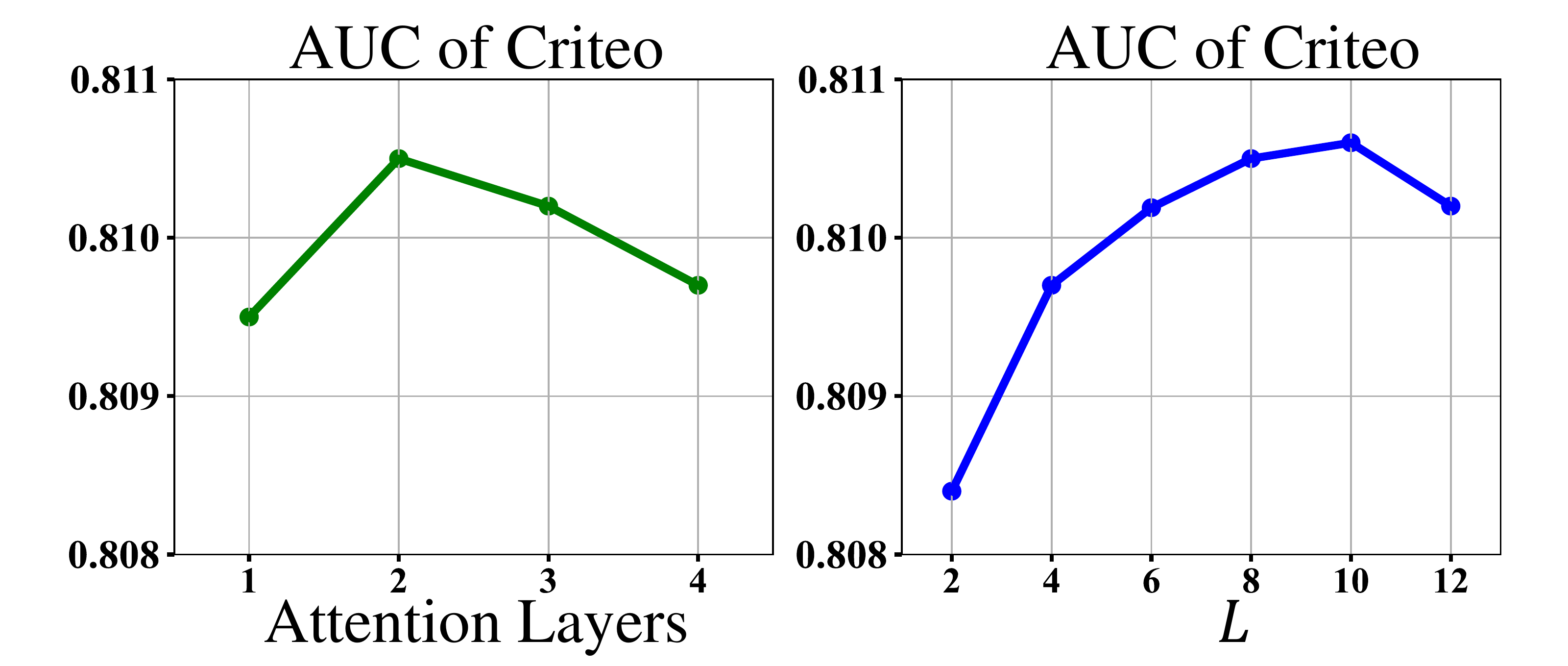}
                \caption{Hyper-Parameter Analysis of DLF}
            \label{fig:line_chart}
            \end{figure}

            To better understand the impact of key architectural hyper-parameters on the performance of our model, we conduct a detailed analysis of two critical factors: the number of attention layers and stacked interaction layers \(L\). The results are visualized in Figure~\ref{fig:line_chart}.
            
            For the number of attention layers, we evaluate the model's performance with 1, 2, 3, and 4 layers. As shown in Figure~\ref{fig:line_chart}, the AUC improves significantly as the number of attention layers increases from 1 to 2, achieving the best performance at 2 layers. However, further increasing the number of attention layers to 3 or 4 results in a decline in AUC. This trend aligns with our expectations: while a moderate number of attention layers enhances the model's ability to effectively fuse explicit and implicit feature interactions, excessive layers may introduce redundancy or overfitting, ultimately degrading the model’s performance. This result highlights the importance of balancing model complexity and feature fusion depth.
            
            For the number of stacked interaction layers \(L\), we test values of 2, 4, 6, 8, 10, and 12. As illustrated in Figure Y, the AUC steadily increases as \(L\) grows, peaking at 8 layers. Beyond this point, the performance plateaus, with 10 layers achieving comparable AUC to 8 layers, and a slight decline observed at 12 layers. This finding suggests that our model is capable of leveraging deeper architectures (up to 8 or 10 layers) to capture high-order feature interactions without suffering from overfitting or gradient vanishing, which are common issues in traditional feature interaction models. In contrast, prior models often reach optimal performance at much shallower depths (e.g., 2 or 3 layers), as their ability to mitigate sparsity issues diminishes with increasing depth.
            
            The observed trends in both attention layers and stacked interaction layers reflect the robustness and scalability of our proposed model. The initial performance improvement with increasing layer depth demonstrates the model's capacity to progressively capture richer feature interactions. More importantly, the ability of our model to maintain strong performance at deeper levels (8 or 10 layers) indicates its effectiveness in alleviating the sparsity challenges often associated with high-order feature interactions. This deeper architecture allows our model to extract more comprehensive and meaningful feature information compared to traditional shallow models, further validating the design of our attention fusion and residual-aware mechanisms.

\section{Conclusion}
In this work, we addressed the challenges of feature interaction modeling in CTR prediction by proposing a novel framework, Dynamic Low-Order-Aware Fusion. The framework was designed to overcome the limitations of existing two-stream architectures, including insufficient information sharing, gradient imbalance, and the difficulty of preserving low-order signals in sparse CTR data. 
We introduced two key components: Residual-Aware Low-Order Interaction Network and Network-Aware Attention Fusion Module. RLI explicitly preserved low-order signals while mitigating redundancy and interference caused by residual connections, ensuring robust propagation of low-order interactions throughout the network. NAF dynamically integrated explicit and implicit representations at each layer through a multi-level attention mechanism, enabling effective collaboration between the two streams and alleviating gradient imbalance during optimization.
Extensive experiments conducted on multiple widely used public datasets demonstrated that DLF consistently achieved state-of-the-art performance in CTR prediction tasks, thereby validating its effectiveness in addressing the inherent limitations of existing models. 
Furthermore, the modular and flexible design of DLF enables it to be seamlessly extended with future advanced feature interaction modules, making it a highly versatile and robust solution for CTR prediction across diverse application scenarios. 
Our work provides a comprehensive and unified solution for explicit-implicit interaction modeling, contributing both novel theoretical insights and significant practical advancements to the rapidly evolving field of CTR prediction.

\begin{acks}
This work was supported by the National Natural Science Foundation of China (U23A20319, 62441239, 62441227, 62472394, 62202443) as well as the Anhui Province Science and Technology Innovation Project (202423k09020011) and Anhui Provincial Science and Technology Major Project (2023z020006).
\end{acks}

\balance


\bibliographystyle{ACM-Reference-Format}
\bibliography{main}


\begin{thebibliography}{79}


\ifx \showCODEN    \undefined \def \showCODEN     #1{\unskip}     \fi
\ifx \showDOI      \undefined \def \showDOI       #1{#1}\fi
\ifx \showISBNx    \undefined \def \showISBNx     #1{\unskip}     \fi
\ifx \showISBNxiii \undefined \def \showISBNxiii  #1{\unskip}     \fi
\ifx \showISSN     \undefined \def \showISSN      #1{\unskip}     \fi
\ifx \showLCCN     \undefined \def \showLCCN      #1{\unskip}     \fi
\ifx \shownote     \undefined \def \shownote      #1{#1}          \fi
\ifx \showarticletitle \undefined \def \showarticletitle #1{#1}   \fi
\ifx \showURL      \undefined \def \showURL       {\relax}        \fi
\providecommand\bibfield[2]{#2}
\providecommand\bibinfo[2]{#2}
\providecommand\natexlab[1]{#1}
\providecommand\showeprint[2][]{arXiv:#2}

\bibitem[Blondel et~al\mbox{.}(2016)]%
        {HOFM}
\bibfield{author}{\bibinfo{person}{Mathieu Blondel}, \bibinfo{person}{Akinori Fujino}, \bibinfo{person}{Naonori Ueda}, {and} \bibinfo{person}{Masakazu Ishihata}.} \bibinfo{year}{2016}\natexlab{}.
\newblock \showarticletitle{Higher-order factorization machines}.
\newblock \bibinfo{journal}{\emph{Advances in neural information processing systems}}  \bibinfo{volume}{29} (\bibinfo{year}{2016}).
\newblock


\bibitem[Chen et~al\mbox{.}(2021)]%
        {EDCN}
\bibfield{author}{\bibinfo{person}{Bo Chen}, \bibinfo{person}{Yichao Wang}, \bibinfo{person}{Zhirong Liu}, \bibinfo{person}{Ruiming Tang}, \bibinfo{person}{Wei Guo}, \bibinfo{person}{Hongkun Zheng}, \bibinfo{person}{Weiwei Yao}, \bibinfo{person}{Muyu Zhang}, {and} \bibinfo{person}{Xiuqiang He}.} \bibinfo{year}{2021}\natexlab{}.
\newblock \showarticletitle{Enhancing explicit and implicit feature interactions via information sharing for parallel deep ctr models}. In \bibinfo{booktitle}{\emph{Proceedings of the 30th ACM international conference on information \& knowledge management}}. \bibinfo{pages}{3757--3766}.
\newblock


\bibitem[Cheng et~al\mbox{.}(2016)]%
        {WideDeep}
\bibfield{author}{\bibinfo{person}{Heng-Tze Cheng}, \bibinfo{person}{Levent Koc}, \bibinfo{person}{Jeremiah Harmsen}, \bibinfo{person}{Tal Shaked}, \bibinfo{person}{Tushar Chandra}, \bibinfo{person}{Hrishi Aradhye}, \bibinfo{person}{Glen Anderson}, \bibinfo{person}{Greg Corrado}, \bibinfo{person}{Wei Chai}, \bibinfo{person}{Mustafa Ispir}, {et~al\mbox{.}}} \bibinfo{year}{2016}\natexlab{}.
\newblock \showarticletitle{Wide \& deep learning for recommender systems}. In \bibinfo{booktitle}{\emph{Proceedings of the 1st workshop on deep learning for recommender systems}}. \bibinfo{pages}{7--10}.
\newblock


\bibitem[Cheng et~al\mbox{.}(2020)]%
        {AFN}
\bibfield{author}{\bibinfo{person}{Weiyu Cheng}, \bibinfo{person}{Yanyan Shen}, {and} \bibinfo{person}{Linpeng Huang}.} \bibinfo{year}{2020}\natexlab{}.
\newblock \showarticletitle{Adaptive factorization network: Learning adaptive-order feature interactions}. In \bibinfo{booktitle}{\emph{Proceedings of the AAAI Conference on Artificial Intelligence}}, Vol.~\bibinfo{volume}{34}. \bibinfo{pages}{3609--3616}.
\newblock


\bibitem[Cheng and Xue(2021)]%
        {SAM}
\bibfield{author}{\bibinfo{person}{Yuan Cheng} {and} \bibinfo{person}{Yanbo Xue}.} \bibinfo{year}{2021}\natexlab{}.
\newblock \showarticletitle{Looking at CTR Prediction Again: Is Attention All You Need?}. In \bibinfo{booktitle}{\emph{Proceedings of the 44th International ACM SIGIR Conference on Research and Development in Information Retrieval}} \emph{(\bibinfo{series}{SIGIR ’21})}. \bibinfo{publisher}{ACM}.
\newblock
\urldef\tempurl%
\url{https://doi.org/10.1145/3404835.3462936}
\showDOI{\tempurl}


\bibitem[Cox(1958)]%
        {LogicR}
\bibfield{author}{\bibinfo{person}{D.~R. Cox}.} \bibinfo{year}{1958}\natexlab{}.
\newblock \showarticletitle{The Regression Analysis of Binary Sequences}.
\newblock \bibinfo{journal}{\emph{Journal of the Royal Statistical Society. Series B (Methodological)}} \bibinfo{volume}{20}, \bibinfo{number}{2} (\bibinfo{year}{1958}), \bibinfo{pages}{215--242}.
\newblock
\showISSN{00359246}
\urldef\tempurl%
\url{http://www.jstor.org/stable/2983890}
\showURL{%
\tempurl}


\bibitem[Deng et~al\mbox{.}(2021a)]%
        {deng2021deeplight}
\bibfield{author}{\bibinfo{person}{Wei Deng}, \bibinfo{person}{Junwei Pan}, \bibinfo{person}{Tian Zhou}, \bibinfo{person}{Deguang Kong}, \bibinfo{person}{Aaron Flores}, {and} \bibinfo{person}{Guang Lin}.} \bibinfo{year}{2021}\natexlab{a}.
\newblock \showarticletitle{Deeplight: Deep lightweight feature interactions for accelerating ctr predictions in ad serving}. In \bibinfo{booktitle}{\emph{Proceedings of the 14th ACM international conference on Web search and data mining}}. \bibinfo{pages}{922--930}.
\newblock


\bibitem[Deng et~al\mbox{.}(2021b)]%
        {DeepLight}
\bibfield{author}{\bibinfo{person}{Wei Deng}, \bibinfo{person}{Junwei Pan}, \bibinfo{person}{Tian Zhou}, \bibinfo{person}{Deguang Kong}, \bibinfo{person}{Aaron Flores}, {and} \bibinfo{person}{Guang Lin}.} \bibinfo{year}{2021}\natexlab{b}.
\newblock \showarticletitle{Deeplight: Deep lightweight feature interactions for accelerating ctr predictions in ad serving}. In \bibinfo{booktitle}{\emph{Proceedings of the 14th ACM international conference on Web search and data mining}}. \bibinfo{pages}{922--930}.
\newblock


\bibitem[Graepel et~al\mbox{.}(2010)]%
        {graepel2010web}
\bibfield{author}{\bibinfo{person}{Thore Graepel}, \bibinfo{person}{Joaquin~Quinonero Candela}, \bibinfo{person}{Thomas Borchert}, {and} \bibinfo{person}{Ralf Herbrich}.} \bibinfo{year}{2010}\natexlab{}.
\newblock \showarticletitle{Web-scale bayesian click-through rate prediction for sponsored search advertising in microsoft's bing search engine}. Omnipress.
\newblock


\bibitem[Gu et~al\mbox{.}(2025)]%
        {gu2025rapid}
\bibfield{author}{\bibinfo{person}{Hongchao Gu}, \bibinfo{person}{Dexun Li}, \bibinfo{person}{Kuicai Dong}, \bibinfo{person}{Hao Zhang}, \bibinfo{person}{Hang Lv}, \bibinfo{person}{Hao Wang}, \bibinfo{person}{Defu Lian}, \bibinfo{person}{Yong Liu}, {and} \bibinfo{person}{Enhong Chen}.} \bibinfo{year}{2025}\natexlab{}.
\newblock \showarticletitle{RAPID: Efficient Retrieval-Augmented Long Text Generation with Writing Planning and Information Discovery}.
\newblock \bibinfo{journal}{\emph{arXiv preprint arXiv:2503.00751}} (\bibinfo{year}{2025}).
\newblock


\bibitem[Guo et~al\mbox{.}(2017)]%
        {DeepFM}
\bibfield{author}{\bibinfo{person}{Huifeng Guo}, \bibinfo{person}{Ruiming Tang}, \bibinfo{person}{Yunming Ye}, \bibinfo{person}{Zhenguo Li}, {and} \bibinfo{person}{Xiuqiang He}.} \bibinfo{year}{2017}\natexlab{}.
\newblock \showarticletitle{DeepFM: a factorization-machine based neural network for CTR prediction}.
\newblock \bibinfo{journal}{\emph{arXiv preprint arXiv:1703.04247}} (\bibinfo{year}{2017}).
\newblock


\bibitem[Guo et~al\mbox{.}(2019)]%
        {OENN}
\bibfield{author}{\bibinfo{person}{Wei Guo}, \bibinfo{person}{Ruiming Tang}, \bibinfo{person}{Huifeng Guo}, \bibinfo{person}{Jianhua Han}, \bibinfo{person}{Wen Yang}, {and} \bibinfo{person}{Yuzhou Zhang}.} \bibinfo{year}{2019}\natexlab{}.
\newblock \showarticletitle{Order-aware embedding neural network for CTR prediction}. In \bibinfo{booktitle}{\emph{Proceedings of the 42nd International ACM SIGIR Conference on Research and Development in Information Retrieval}}. \bibinfo{pages}{1121--1124}.
\newblock


\bibitem[Guo et~al\mbox{.}(2024)]%
        {guo2024scaling}
\bibfield{author}{\bibinfo{person}{Wei Guo}, \bibinfo{person}{Hao Wang}, \bibinfo{person}{Luankang Zhang}, \bibinfo{person}{Jin~Yao Chin}, \bibinfo{person}{Zhongzhou Liu}, \bibinfo{person}{Kai Cheng}, \bibinfo{person}{Qiushi Pan}, \bibinfo{person}{Yi~Quan Lee}, \bibinfo{person}{Wanqi Xue}, \bibinfo{person}{Tingjia Shen}, {et~al\mbox{.}}} \bibinfo{year}{2024}\natexlab{}.
\newblock \showarticletitle{Scaling New Frontiers: Insights into Large Recommendation Models}.
\newblock \bibinfo{journal}{\emph{arXiv preprint arXiv:2412.00714}} (\bibinfo{year}{2024}).
\newblock


\bibitem[Han et~al\mbox{.}(2024)]%
        {han2024efficient}
\bibfield{author}{\bibinfo{person}{Yongqiang Han}, \bibinfo{person}{Hao Wang}, \bibinfo{person}{Kefan Wang}, \bibinfo{person}{Likang Wu}, \bibinfo{person}{Zhi Li}, \bibinfo{person}{Wei Guo}, \bibinfo{person}{Yong Liu}, \bibinfo{person}{Defu Lian}, {and} \bibinfo{person}{Enhong Chen}.} \bibinfo{year}{2024}\natexlab{}.
\newblock \showarticletitle{Efficient noise-decoupling for multi-behavior sequential recommendation}. In \bibinfo{booktitle}{\emph{Proceedings of the ACM Web Conference 2024}}. \bibinfo{pages}{3297--3306}.
\newblock


\bibitem[Han et~al\mbox{.}(2023)]%
        {han2023guesr}
\bibfield{author}{\bibinfo{person}{Yongqiang Han}, \bibinfo{person}{Likang Wu}, \bibinfo{person}{Hao Wang}, \bibinfo{person}{Guifeng Wang}, \bibinfo{person}{Mengdi Zhang}, \bibinfo{person}{Zhi Li}, \bibinfo{person}{Defu Lian}, {and} \bibinfo{person}{Enhong Chen}.} \bibinfo{year}{2023}\natexlab{}.
\newblock \showarticletitle{Guesr: A global unsupervised data-enhancement with bucket-cluster sampling for sequential recommendation}. In \bibinfo{booktitle}{\emph{International conference on database systems for advanced applications}}. Springer, \bibinfo{pages}{286--296}.
\newblock


\bibitem[He and Chua(2017)]%
        {NFM}
\bibfield{author}{\bibinfo{person}{Xiangnan He} {and} \bibinfo{person}{Tat-Seng Chua}.} \bibinfo{year}{2017}\natexlab{}.
\newblock \bibinfo{title}{Neural Factorization Machines for Sparse Predictive Analytics}.
\newblock
\newblock
\showeprint[arxiv]{1708.05027}~[cs.LG]
\urldef\tempurl%
\url{https://arxiv.org/abs/1708.05027}
\showURL{%
\tempurl}


\bibitem[He et~al\mbox{.}(2014)]%
        {he2014practical}
\bibfield{author}{\bibinfo{person}{Xinran He}, \bibinfo{person}{Junfeng Pan}, \bibinfo{person}{Ou Jin}, \bibinfo{person}{Tianbing Xu}, \bibinfo{person}{Bo Liu}, \bibinfo{person}{Tao Xu}, \bibinfo{person}{Yanxin Shi}, \bibinfo{person}{Antoine Atallah}, \bibinfo{person}{Ralf Herbrich}, \bibinfo{person}{Stuart Bowers}, {et~al\mbox{.}}} \bibinfo{year}{2014}\natexlab{}.
\newblock \showarticletitle{Practical lessons from predicting clicks on ads at facebook}. In \bibinfo{booktitle}{\emph{Proceedings of the eighth international workshop on data mining for online advertising}}. \bibinfo{pages}{1--9}.
\newblock


\bibitem[Hu et~al\mbox{.}(2019)]%
        {hu2019dense}
\bibfield{author}{\bibinfo{person}{Di Hu}, \bibinfo{person}{Chengze Wang}, \bibinfo{person}{Feiping Nie}, {and} \bibinfo{person}{Xuelong Li}.} \bibinfo{year}{2019}\natexlab{}.
\newblock \showarticletitle{Dense multimodal fusion for hierarchically joint representation}. In \bibinfo{booktitle}{\emph{ICASSP 2019-2019 IEEE International Conference on Acoustics, Speech and Signal Processing (ICASSP)}}. IEEE, \bibinfo{pages}{3941--3945}.
\newblock


\bibitem[Hu et~al\mbox{.}(2018)]%
        {SENet}
\bibfield{author}{\bibinfo{person}{Jie Hu}, \bibinfo{person}{Li Shen}, {and} \bibinfo{person}{Gang Sun}.} \bibinfo{year}{2018}\natexlab{}.
\newblock \showarticletitle{Squeeze-and-excitation networks}. In \bibinfo{booktitle}{\emph{Proceedings of the IEEE conference on computer vision and pattern recognition}}. \bibinfo{pages}{7132--7141}.
\newblock


\bibitem[Huang and Ling(2005)]%
        {AUC}
\bibfield{author}{\bibinfo{person}{Jin Huang} {and} \bibinfo{person}{Charles~X Ling}.} \bibinfo{year}{2005}\natexlab{}.
\newblock \showarticletitle{Using AUC and accuracy in evaluating learning algorithms}.
\newblock \bibinfo{journal}{\emph{IEEE Transactions on knowledge and Data Engineering}} \bibinfo{volume}{17}, \bibinfo{number}{3} (\bibinfo{year}{2005}), \bibinfo{pages}{299--310}.
\newblock


\bibitem[Huang et~al\mbox{.}(2019)]%
        {FiBiNET}
\bibfield{author}{\bibinfo{person}{Tongwen Huang}, \bibinfo{person}{Zhiqi Zhang}, {and} \bibinfo{person}{Junlin Zhang}.} \bibinfo{year}{2019}\natexlab{}.
\newblock \showarticletitle{FiBiNET: combining feature importance and bilinear feature interaction for click-through rate prediction}. In \bibinfo{booktitle}{\emph{Proceedings of the 13th ACM conference on recommender systems}}. \bibinfo{pages}{169--177}.
\newblock


\bibitem[Huang et~al\mbox{.}(2024)]%
        {huang2024chemeval}
\bibfield{author}{\bibinfo{person}{Yuqing Huang}, \bibinfo{person}{Rongyang Zhang}, \bibinfo{person}{Xuesong He}, \bibinfo{person}{Xuyang Zhi}, \bibinfo{person}{Hao Wang}, \bibinfo{person}{Xin Li}, \bibinfo{person}{Feiyang Xu}, \bibinfo{person}{Deguang Liu}, \bibinfo{person}{Huadong Liang}, \bibinfo{person}{Yi Li}, {et~al\mbox{.}}} \bibinfo{year}{2024}\natexlab{}.
\newblock \showarticletitle{ChemEval: A Comprehensive Multi-Level Chemical Evaluation for Large Language Models}.
\newblock \bibinfo{journal}{\emph{arXiv preprint arXiv:2409.13989}} (\bibinfo{year}{2024}).
\newblock


\bibitem[Kingma and Ba(2017)]%
        {adam}
\bibfield{author}{\bibinfo{person}{Diederik~P. Kingma} {and} \bibinfo{person}{Jimmy Ba}.} \bibinfo{year}{2017}\natexlab{}.
\newblock \bibinfo{title}{Adam: A Method for Stochastic Optimization}.
\newblock
\newblock
\showeprint[arxiv]{1412.6980}~[cs.LG]
\urldef\tempurl%
\url{https://arxiv.org/abs/1412.6980}
\showURL{%
\tempurl}


\bibitem[Li et~al\mbox{.}(2024)]%
        {DCNv3}
\bibfield{author}{\bibinfo{person}{Honghao Li}, \bibinfo{person}{Yiwen Zhang}, \bibinfo{person}{Yi Zhang}, \bibinfo{person}{Hanwei Li}, \bibinfo{person}{Lei Sang}, {and} \bibinfo{person}{Jieming Zhu}.} \bibinfo{year}{2024}\natexlab{}.
\newblock \showarticletitle{DCNv3: Towards Next Generation Deep Cross Network for CTR Prediction}.
\newblock \bibinfo{journal}{\emph{arXiv preprint arXiv:2407.13349}} (\bibinfo{year}{2024}).
\newblock


\bibitem[Lian et~al\mbox{.}(2018)]%
        {xDeepFM}
\bibfield{author}{\bibinfo{person}{Jianxun Lian}, \bibinfo{person}{Xiaohuan Zhou}, \bibinfo{person}{Fuzheng Zhang}, \bibinfo{person}{Zhongxia Chen}, \bibinfo{person}{Xing Xie}, {and} \bibinfo{person}{Guangzhong Sun}.} \bibinfo{year}{2018}\natexlab{}.
\newblock \showarticletitle{xDeepFM: Combining Explicit and Implicit Feature Interactions for Recommender Systems}. In \bibinfo{booktitle}{\emph{Proceedings of the 24th ACM SIGKDD International Conference on Knowledge Discovery \&amp; Data Mining}} \emph{(\bibinfo{series}{KDD ’18})}. \bibinfo{publisher}{ACM}.
\newblock
\urldef\tempurl%
\url{https://doi.org/10.1145/3219819.3220023}
\showDOI{\tempurl}


\bibitem[Liao et~al\mbox{.}(2023)]%
        {FusionChallenge3}
\bibfield{author}{\bibinfo{person}{Chao Liao}, \bibinfo{person}{Jianchao Tan}, \bibinfo{person}{Jiyuan Jia}, \bibinfo{person}{Yi Guo}, {and} \bibinfo{person}{Chengru Song}.} \bibinfo{year}{2023}\natexlab{}.
\newblock \showarticletitle{MaskFusion: Feature Augmentation for Click-Through Rate Prediction via Input-adaptive Mask Fusion}. In \bibinfo{booktitle}{\emph{The Eleventh International Conference on Learning Representations}}.
\newblock


\bibitem[Liu et~al\mbox{.}(2020a)]%
        {AutoGroup}
\bibfield{author}{\bibinfo{person}{Bin Liu}, \bibinfo{person}{Niannan Xue}, \bibinfo{person}{Huifeng Guo}, \bibinfo{person}{Ruiming Tang}, \bibinfo{person}{Stefanos Zafeiriou}, \bibinfo{person}{Xiuqiang He}, {and} \bibinfo{person}{Zhenguo Li}.} \bibinfo{year}{2020}\natexlab{a}.
\newblock \showarticletitle{AutoGroup: Automatic feature grouping for modelling explicit high-order feature interactions in CTR prediction}. In \bibinfo{booktitle}{\emph{Proceedings of the 43rd international ACM SIGIR conference on research and development in information retrieval}}. \bibinfo{pages}{199--208}.
\newblock


\bibitem[Liu et~al\mbox{.}(2020b)]%
        {liu2020autofis}
\bibfield{author}{\bibinfo{person}{Bin Liu}, \bibinfo{person}{Chenxu Zhu}, \bibinfo{person}{Guilin Li}, \bibinfo{person}{Weinan Zhang}, \bibinfo{person}{Jincai Lai}, \bibinfo{person}{Ruiming Tang}, \bibinfo{person}{Xiuqiang He}, \bibinfo{person}{Zhenguo Li}, {and} \bibinfo{person}{Yong Yu}.} \bibinfo{year}{2020}\natexlab{b}.
\newblock \showarticletitle{Autofis: Automatic feature interaction selection in factorization models for click-through rate prediction}. In \bibinfo{booktitle}{\emph{proceedings of the 26th ACM SIGKDD international conference on knowledge discovery \& data mining}}. \bibinfo{pages}{2636--2645}.
\newblock


\bibitem[Liu et~al\mbox{.}(2020c)]%
        {AutoFis}
\bibfield{author}{\bibinfo{person}{Bin Liu}, \bibinfo{person}{Chenxu Zhu}, \bibinfo{person}{Guilin Li}, \bibinfo{person}{Weinan Zhang}, \bibinfo{person}{Jincai Lai}, \bibinfo{person}{Ruiming Tang}, \bibinfo{person}{Xiuqiang He}, \bibinfo{person}{Zhenguo Li}, {and} \bibinfo{person}{Yong Yu}.} \bibinfo{year}{2020}\natexlab{c}.
\newblock \showarticletitle{Autofis: Automatic feature interaction selection in factorization models for click-through rate prediction}. In \bibinfo{booktitle}{\emph{proceedings of the 26th ACM SIGKDD international conference on knowledge discovery \& data mining}}. \bibinfo{pages}{2636--2645}.
\newblock


\bibitem[Liu et~al\mbox{.}(2023)]%
        {liu2023user}
\bibfield{author}{\bibinfo{person}{Weiwen Liu}, \bibinfo{person}{Wei Guo}, \bibinfo{person}{Yong Liu}, \bibinfo{person}{Ruiming Tang}, {and} \bibinfo{person}{Hao Wang}.} \bibinfo{year}{2023}\natexlab{}.
\newblock \showarticletitle{User Behavior Modeling with Deep Learning for Recommendation: Recent Advances}. In \bibinfo{booktitle}{\emph{Proceedings of the 17th ACM Conference on Recommender Systems}}. \bibinfo{pages}{1286--1287}.
\newblock


\bibitem[Lu et~al\mbox{.}(2021)]%
        {DIFM}
\bibfield{author}{\bibinfo{person}{Wantong Lu}, \bibinfo{person}{Yantao Yu}, \bibinfo{person}{Yongzhe Chang}, \bibinfo{person}{Zhen Wang}, \bibinfo{person}{Chenhui Li}, {and} \bibinfo{person}{Bo Yuan}.} \bibinfo{year}{2021}\natexlab{}.
\newblock \showarticletitle{A dual input-aware factorization machine for CTR prediction}. In \bibinfo{booktitle}{\emph{Proceedings of the twenty-ninth international conference on international joint conferences on artificial intelligence}}. \bibinfo{pages}{3139--3145}.
\newblock


\bibitem[Ma et~al\mbox{.}(2018b)]%
        {MMoE}
\bibfield{author}{\bibinfo{person}{Jiaqi Ma}, \bibinfo{person}{Zhe Zhao}, \bibinfo{person}{Xinyang Yi}, \bibinfo{person}{Jilin Chen}, \bibinfo{person}{Lichan Hong}, {and} \bibinfo{person}{Ed~H Chi}.} \bibinfo{year}{2018}\natexlab{b}.
\newblock \showarticletitle{Modeling task relationships in multi-task learning with multi-gate mixture-of-experts}. In \bibinfo{booktitle}{\emph{Proceedings of the 24th ACM SIGKDD international conference on knowledge discovery \& data mining}}. \bibinfo{pages}{1930--1939}.
\newblock


\bibitem[Ma et~al\mbox{.}(2018a)]%
        {ESMM}
\bibfield{author}{\bibinfo{person}{Xiao Ma}, \bibinfo{person}{Liqin Zhao}, \bibinfo{person}{Guan Huang}, \bibinfo{person}{Zhi Wang}, \bibinfo{person}{Zelin Hu}, \bibinfo{person}{Xiaoqiang Zhu}, {and} \bibinfo{person}{Kun Gai}.} \bibinfo{year}{2018}\natexlab{a}.
\newblock \showarticletitle{Entire space multi-task model: An effective approach for estimating post-click conversion rate}. In \bibinfo{booktitle}{\emph{The 41st International ACM SIGIR Conference on Research \& Development in Information Retrieval}}. \bibinfo{pages}{1137--1140}.
\newblock


\bibitem[Mao et~al\mbox{.}(2023)]%
        {FinalMLP}
\bibfield{author}{\bibinfo{person}{Kelong Mao}, \bibinfo{person}{Jieming Zhu}, \bibinfo{person}{Liangcai Su}, \bibinfo{person}{Guohao Cai}, \bibinfo{person}{Yuru Li}, {and} \bibinfo{person}{Zhenhua Dong}.} \bibinfo{year}{2023}\natexlab{}.
\newblock \showarticletitle{FinalMLP: an enhanced two-stream MLP model for CTR prediction}. In \bibinfo{booktitle}{\emph{Proceedings of the AAAI Conference on Artificial Intelligence}}, Vol.~\bibinfo{volume}{37}. \bibinfo{pages}{4552--4560}.
\newblock


\bibitem[Ouyang et~al\mbox{.}(2019)]%
        {DeepMCP}
\bibfield{author}{\bibinfo{person}{Wentao Ouyang}, \bibinfo{person}{Xiuwu Zhang}, \bibinfo{person}{Shukui Ren}, \bibinfo{person}{Chao Qi}, \bibinfo{person}{Zhaojie Liu}, {and} \bibinfo{person}{Yanlong Du}.} \bibinfo{year}{2019}\natexlab{}.
\newblock \showarticletitle{Representation learning-assisted click-through rate prediction}.
\newblock \bibinfo{journal}{\emph{arXiv preprint arXiv:1906.04365}} (\bibinfo{year}{2019}).
\newblock


\bibitem[Pan et~al\mbox{.}(2018)]%
        {FwFM}
\bibfield{author}{\bibinfo{person}{Junwei Pan}, \bibinfo{person}{Jian Xu}, \bibinfo{person}{Alfonso~Lobos Ruiz}, \bibinfo{person}{Wenliang Zhao}, \bibinfo{person}{Shengjun Pan}, \bibinfo{person}{Yu Sun}, {and} \bibinfo{person}{Quan Lu}.} \bibinfo{year}{2018}\natexlab{}.
\newblock \showarticletitle{Field-weighted factorization machines for click-through rate prediction in display advertising}. In \bibinfo{booktitle}{\emph{Proceedings of the 2018 world wide web conference}}. \bibinfo{pages}{1349--1357}.
\newblock


\bibitem[Qu et~al\mbox{.}(2016)]%
        {PNN}
\bibfield{author}{\bibinfo{person}{Yanru Qu}, \bibinfo{person}{Han Cai}, \bibinfo{person}{Kan Ren}, \bibinfo{person}{Weinan Zhang}, \bibinfo{person}{Yong Yu}, \bibinfo{person}{Ying Wen}, {and} \bibinfo{person}{Jun Wang}.} \bibinfo{year}{2016}\natexlab{}.
\newblock \showarticletitle{Product-based neural networks for user response prediction}. In \bibinfo{booktitle}{\emph{2016 IEEE 16th international conference on data mining (ICDM)}}. IEEE, \bibinfo{pages}{1149--1154}.
\newblock


\bibitem[Rendle(2010)]%
        {FM}
\bibfield{author}{\bibinfo{person}{Steffen Rendle}.} \bibinfo{year}{2010}\natexlab{}.
\newblock \showarticletitle{Factorization machines}. In \bibinfo{booktitle}{\emph{2010 IEEE International conference on data mining}}. IEEE, \bibinfo{pages}{995--1000}.
\newblock


\bibitem[Rendle et~al\mbox{.}(2020)]%
        {rendle2020neural}
\bibfield{author}{\bibinfo{person}{Steffen Rendle}, \bibinfo{person}{Walid Krichene}, \bibinfo{person}{Li Zhang}, {and} \bibinfo{person}{John Anderson}.} \bibinfo{year}{2020}\natexlab{}.
\newblock \showarticletitle{Neural collaborative filtering vs. matrix factorization revisited}. In \bibinfo{booktitle}{\emph{Proceedings of the 14th ACM Conference on Recommender Systems}}. \bibinfo{pages}{240--248}.
\newblock


\bibitem[Richardson et~al\mbox{.}(2007)]%
        {LR}
\bibfield{author}{\bibinfo{person}{Matthew Richardson}, \bibinfo{person}{Ewa Dominowska}, {and} \bibinfo{person}{Robert Ragno}.} \bibinfo{year}{2007}\natexlab{}.
\newblock \showarticletitle{Predicting clicks: estimating the click-through rate for new ads}. In \bibinfo{booktitle}{\emph{Proceedings of the 16th international conference on World Wide Web}}. \bibinfo{pages}{521--530}.
\newblock


\bibitem[Sang et~al\mbox{.}(2024)]%
        {FusionChallenge1}
\bibfield{author}{\bibinfo{person}{Lei Sang}, \bibinfo{person}{Qiuze Ru}, \bibinfo{person}{Honghao Li}, \bibinfo{person}{Yiwen Zhang}, \bibinfo{person}{Qian Cao}, {and} \bibinfo{person}{Xindong Wu}.} \bibinfo{year}{2024}\natexlab{}.
\newblock \showarticletitle{Feature Interaction Fusion Self-Distillation Network For CTR Prediction}.
\newblock \bibinfo{journal}{\emph{arXiv preprint arXiv:2411.07508}} (\bibinfo{year}{2024}).
\newblock


\bibitem[Shen et~al\mbox{.}(2024a)]%
        {shen2024optimizing}
\bibfield{author}{\bibinfo{person}{Tingjia Shen}, \bibinfo{person}{Hao Wang}, \bibinfo{person}{Chuhan Wu}, \bibinfo{person}{Jin~Yao Chin}, \bibinfo{person}{Wei Guo}, \bibinfo{person}{Yong Liu}, \bibinfo{person}{Huifeng Guo}, \bibinfo{person}{Defu Lian}, \bibinfo{person}{Ruiming Tang}, {and} \bibinfo{person}{Enhong Chen}.} \bibinfo{year}{2024}\natexlab{a}.
\newblock \showarticletitle{Optimizing Sequential Recommendation Models with Scaling Laws and Approximate Entropy}.
\newblock \bibinfo{journal}{\emph{arXiv preprint arXiv:2412.00430}} (\bibinfo{year}{2024}).
\newblock


\bibitem[Shen et~al\mbox{.}(2024b)]%
        {shen2024exploring}
\bibfield{author}{\bibinfo{person}{Tingjia Shen}, \bibinfo{person}{Hao Wang}, \bibinfo{person}{Jiaqing Zhang}, \bibinfo{person}{Sirui Zhao}, \bibinfo{person}{Liangyue Li}, \bibinfo{person}{Zulong Chen}, \bibinfo{person}{Defu Lian}, {and} \bibinfo{person}{Enhong Chen}.} \bibinfo{year}{2024}\natexlab{b}.
\newblock \showarticletitle{Exploring user retrieval integration towards large language models for cross-domain sequential recommendation}.
\newblock \bibinfo{journal}{\emph{arXiv preprint arXiv:2406.03085}} (\bibinfo{year}{2024}).
\newblock


\bibitem[Song et~al\mbox{.}(2020)]%
        {AutoCTR}
\bibfield{author}{\bibinfo{person}{Qingquan Song}, \bibinfo{person}{Dehua Cheng}, \bibinfo{person}{Hanning Zhou}, \bibinfo{person}{Jiyan Yang}, \bibinfo{person}{Yuandong Tian}, {and} \bibinfo{person}{Xia Hu}.} \bibinfo{year}{2020}\natexlab{}.
\newblock \showarticletitle{Towards automated neural interaction discovery for click-through rate prediction}. In \bibinfo{booktitle}{\emph{Proceedings of the 26th ACM SIGKDD International Conference on Knowledge Discovery \& Data Mining}}. \bibinfo{pages}{945--955}.
\newblock


\bibitem[Song et~al\mbox{.}(2019)]%
        {AutoInt}
\bibfield{author}{\bibinfo{person}{Weiping Song}, \bibinfo{person}{Chence Shi}, \bibinfo{person}{Zhiping Xiao}, \bibinfo{person}{Zhijian Duan}, \bibinfo{person}{Yewen Xu}, \bibinfo{person}{Ming Zhang}, {and} \bibinfo{person}{Jian Tang}.} \bibinfo{year}{2019}\natexlab{}.
\newblock \showarticletitle{Autoint: Automatic feature interaction learning via self-attentive neural networks}. In \bibinfo{booktitle}{\emph{Proceedings of the 28th ACM international conference on information and knowledge management}}. \bibinfo{pages}{1161--1170}.
\newblock


\bibitem[Tong et~al\mbox{.}(2024)]%
        {tong2024mdap}
\bibfield{author}{\bibinfo{person}{Junxiong Tong}, \bibinfo{person}{Mingjia Yin}, \bibinfo{person}{Hao Wang}, \bibinfo{person}{Qiushi Pan}, \bibinfo{person}{Defu Lian}, {and} \bibinfo{person}{Enhong Chen}.} \bibinfo{year}{2024}\natexlab{}.
\newblock \showarticletitle{MDAP: A Multi-view Disentangled and Adaptive Preference Learning Framework for Cross-Domain Recommendation}. In \bibinfo{booktitle}{\emph{International Conference on Web Information Systems Engineering}}. Springer, \bibinfo{pages}{164--178}.
\newblock


\bibitem[Wang et~al\mbox{.}(2023)]%
        {GDCN}
\bibfield{author}{\bibinfo{person}{Fangye Wang}, \bibinfo{person}{Hansu Gu}, \bibinfo{person}{Dongsheng Li}, \bibinfo{person}{Tun Lu}, \bibinfo{person}{Peng Zhang}, {and} \bibinfo{person}{Ning Gu}.} \bibinfo{year}{2023}\natexlab{}.
\newblock \showarticletitle{Towards deeper, lighter and interpretable cross network for ctr prediction}. In \bibinfo{booktitle}{\emph{Proceedings of the 32nd ACM International Conference on Information and Knowledge Management}}. \bibinfo{pages}{2523--2533}.
\newblock


\bibitem[Wang et~al\mbox{.}(2025a)]%
        {wang2025generative}
\bibfield{author}{\bibinfo{person}{Hao Wang}, \bibinfo{person}{Wei Guo}, \bibinfo{person}{Luankang Zhang}, \bibinfo{person}{Jin~Yao Chin}, \bibinfo{person}{Yufei Ye}, \bibinfo{person}{Huifeng Guo}, \bibinfo{person}{Yong Liu}, \bibinfo{person}{Defu Lian}, \bibinfo{person}{Ruiming Tang}, {and} \bibinfo{person}{Enhong Chen}.} \bibinfo{year}{2025}\natexlab{a}.
\newblock \showarticletitle{Generative Large Recommendation Models: Emerging Trends in LLMs for Recommendation}.
\newblock \bibinfo{journal}{\emph{arXiv preprint arXiv:2502.13783}} (\bibinfo{year}{2025}).
\newblock


\bibitem[Wang et~al\mbox{.}(2024)]%
        {wang2024denoising}
\bibfield{author}{\bibinfo{person}{Hao Wang}, \bibinfo{person}{Yongqiang Han}, \bibinfo{person}{Kefan Wang}, \bibinfo{person}{Kai Cheng}, \bibinfo{person}{Zhen Wang}, \bibinfo{person}{Wei Guo}, \bibinfo{person}{Yong Liu}, \bibinfo{person}{Defu Lian}, {and} \bibinfo{person}{Enhong Chen}.} \bibinfo{year}{2024}\natexlab{}.
\newblock \showarticletitle{Denoising Pre-Training and Customized Prompt Learning for Efficient Multi-Behavior Sequential Recommendation}.
\newblock \bibinfo{journal}{\emph{arXiv preprint arXiv:2408.11372}} (\bibinfo{year}{2024}).
\newblock


\bibitem[Wang et~al\mbox{.}(2021a)]%
        {wang2021decoupled}
\bibfield{author}{\bibinfo{person}{Hao Wang}, \bibinfo{person}{Defu Lian}, \bibinfo{person}{Hanghang Tong}, \bibinfo{person}{Qi Liu}, \bibinfo{person}{Zhenya Huang}, {and} \bibinfo{person}{Enhong Chen}.} \bibinfo{year}{2021}\natexlab{a}.
\newblock \showarticletitle{Decoupled representation learning for attributed networks}.
\newblock \bibinfo{journal}{\emph{IEEE Transactions on Knowledge and Data Engineering}} \bibinfo{volume}{35}, \bibinfo{number}{3} (\bibinfo{year}{2021}), \bibinfo{pages}{2430--2444}.
\newblock


\bibitem[Wang et~al\mbox{.}(2021b)]%
        {wang2021hypersorec}
\bibfield{author}{\bibinfo{person}{Hao Wang}, \bibinfo{person}{Defu Lian}, \bibinfo{person}{Hanghang Tong}, \bibinfo{person}{Qi Liu}, \bibinfo{person}{Zhenya Huang}, {and} \bibinfo{person}{Enhong Chen}.} \bibinfo{year}{2021}\natexlab{b}.
\newblock \showarticletitle{Hypersorec: Exploiting hyperbolic user and item representations with multiple aspects for social-aware recommendation}.
\newblock \bibinfo{journal}{\emph{ACM Transactions on Information Systems (TOIS)}} \bibinfo{volume}{40}, \bibinfo{number}{2} (\bibinfo{year}{2021}), \bibinfo{pages}{1--28}.
\newblock


\bibitem[Wang et~al\mbox{.}(2019b)]%
        {wang2019mcne}
\bibfield{author}{\bibinfo{person}{Hao Wang}, \bibinfo{person}{Tong Xu}, \bibinfo{person}{Qi Liu}, \bibinfo{person}{Defu Lian}, \bibinfo{person}{Enhong Chen}, \bibinfo{person}{Dongfang Du}, \bibinfo{person}{Han Wu}, {and} \bibinfo{person}{Wen Su}.} \bibinfo{year}{2019}\natexlab{b}.
\newblock \showarticletitle{MCNE: An end-to-end framework for learning multiple conditional network representations of social network}. In \bibinfo{booktitle}{\emph{Proceedings of the 25th ACM SIGKDD international conference on knowledge discovery \& data mining}}. \bibinfo{pages}{1064--1072}.
\newblock


\bibitem[Wang et~al\mbox{.}(2025c)]%
        {wang2025mf}
\bibfield{author}{\bibinfo{person}{Hao Wang}, \bibinfo{person}{Mingjia Yin}, \bibinfo{person}{Luankang Zhang}, \bibinfo{person}{Sirui Zhao}, {and} \bibinfo{person}{Enhong Chen}.} \bibinfo{year}{2025}\natexlab{c}.
\newblock \showarticletitle{MF-GSLAE: A Multi-Factor User Representation Pre-training Framework for Dual-Target Cross-Domain Recommendation}.
\newblock \bibinfo{journal}{\emph{ACM Transactions on Information Systems}} \bibinfo{volume}{43}, \bibinfo{number}{2} (\bibinfo{year}{2025}), \bibinfo{pages}{1--28}.
\newblock


\bibitem[Wang et~al\mbox{.}(2025b)]%
        {wang2025universal}
\bibfield{author}{\bibinfo{person}{Kefan Wang}, \bibinfo{person}{Hao Wang}, \bibinfo{person}{Kenan Song}, \bibinfo{person}{Wei Guo}, \bibinfo{person}{Kai Cheng}, \bibinfo{person}{Zhi Li}, \bibinfo{person}{Yong Liu}, \bibinfo{person}{Defu Lian}, {and} \bibinfo{person}{Enhong Chen}.} \bibinfo{year}{2025}\natexlab{b}.
\newblock \showarticletitle{A Universal Framework for Compressing Embeddings in CTR Prediction}.
\newblock \bibinfo{journal}{\emph{arXiv preprint arXiv:2502.15355}} (\bibinfo{year}{2025}).
\newblock


\bibitem[Wang et~al\mbox{.}(2017)]%
        {DCN}
\bibfield{author}{\bibinfo{person}{Ruoxi Wang}, \bibinfo{person}{Bin Fu}, \bibinfo{person}{Gang Fu}, {and} \bibinfo{person}{Mingliang Wang}.} \bibinfo{year}{2017}\natexlab{}.
\newblock \showarticletitle{Deep \& cross network for ad click predictions}.
\newblock In \bibinfo{booktitle}{\emph{Proceedings of the ADKDD'17}}. \bibinfo{pages}{1--7}.
\newblock


\bibitem[Wang et~al\mbox{.}(2021c)]%
        {DCNv2}
\bibfield{author}{\bibinfo{person}{Ruoxi Wang}, \bibinfo{person}{Rakesh Shivanna}, \bibinfo{person}{Derek Cheng}, \bibinfo{person}{Sagar Jain}, \bibinfo{person}{Dong Lin}, \bibinfo{person}{Lichan Hong}, {and} \bibinfo{person}{Ed Chi}.} \bibinfo{year}{2021}\natexlab{c}.
\newblock \showarticletitle{Dcn v2: Improved deep \& cross network and practical lessons for web-scale learning to rank systems}. In \bibinfo{booktitle}{\emph{Proceedings of the web conference 2021}}. \bibinfo{pages}{1785--1797}.
\newblock


\bibitem[Wang et~al\mbox{.}(2019a)]%
        {NGCF}
\bibfield{author}{\bibinfo{person}{Xiang Wang}, \bibinfo{person}{Xiangnan He}, \bibinfo{person}{Meng Wang}, \bibinfo{person}{Fuli Feng}, {and} \bibinfo{person}{Tat-Seng Chua}.} \bibinfo{year}{2019}\natexlab{a}.
\newblock \showarticletitle{Neural graph collaborative filtering}. In \bibinfo{booktitle}{\emph{Proceedings of the 42nd international ACM SIGIR conference on Research and development in Information Retrieval}}. \bibinfo{pages}{165--174}.
\newblock


\bibitem[Wu et~al\mbox{.}(2024)]%
        {wu2024survey}
\bibfield{author}{\bibinfo{person}{Likang Wu}, \bibinfo{person}{Zhi Zheng}, \bibinfo{person}{Zhaopeng Qiu}, \bibinfo{person}{Hao Wang}, \bibinfo{person}{Hongchao Gu}, \bibinfo{person}{Tingjia Shen}, \bibinfo{person}{Chuan Qin}, \bibinfo{person}{Chen Zhu}, \bibinfo{person}{Hengshu Zhu}, \bibinfo{person}{Qi Liu}, {et~al\mbox{.}}} \bibinfo{year}{2024}\natexlab{}.
\newblock \showarticletitle{A survey on large language models for recommendation}.
\newblock \bibinfo{journal}{\emph{World Wide Web}} \bibinfo{volume}{27}, \bibinfo{number}{5} (\bibinfo{year}{2024}), \bibinfo{pages}{60}.
\newblock


\bibitem[Wu et~al\mbox{.}(2021)]%
        {GraphFM}
\bibfield{author}{\bibinfo{person}{Shu Wu}, \bibinfo{person}{Zekun Li}, \bibinfo{person}{Yunyue Su}, \bibinfo{person}{Zeyu Cui}, \bibinfo{person}{Xiaoyu Zhang}, {and} \bibinfo{person}{Liang Wang}.} \bibinfo{year}{2021}\natexlab{}.
\newblock \showarticletitle{GraphFM: Graph factorization machines for feature interaction modeling}.
\newblock \bibinfo{journal}{\emph{arXiv preprint arXiv:2105.11866}} (\bibinfo{year}{2021}).
\newblock


\bibitem[Xiao et~al\mbox{.}(2017)]%
        {AFM}
\bibfield{author}{\bibinfo{person}{Jun Xiao}, \bibinfo{person}{Hao Ye}, \bibinfo{person}{Xiangnan He}, \bibinfo{person}{Hanwang Zhang}, \bibinfo{person}{Fei Wu}, {and} \bibinfo{person}{Tat-Seng Chua}.} \bibinfo{year}{2017}\natexlab{}.
\newblock \bibinfo{title}{Attentional Factorization Machines: Learning the Weight of Feature Interactions via Attention Networks}.
\newblock
\newblock
\showeprint[arxiv]{1708.04617}~[cs.LG]
\urldef\tempurl%
\url{https://arxiv.org/abs/1708.04617}
\showURL{%
\tempurl}


\bibitem[Xie et~al\mbox{.}(2024a)]%
        {xie2024breaking}
\bibfield{author}{\bibinfo{person}{Wenjia Xie}, \bibinfo{person}{Hao Wang}, \bibinfo{person}{Luankang Zhang}, \bibinfo{person}{Rui Zhou}, \bibinfo{person}{Defu Lian}, {and} \bibinfo{person}{Enhong Chen}.} \bibinfo{year}{2024}\natexlab{a}.
\newblock \showarticletitle{Breaking determinism: Fuzzy modeling of sequential recommendation using discrete state space diffusion model}.
\newblock \bibinfo{journal}{\emph{Advances in Neural Information Processing Systems}}  \bibinfo{volume}{37} (\bibinfo{year}{2024}), \bibinfo{pages}{22720--22744}.
\newblock


\bibitem[Xie et~al\mbox{.}(2024b)]%
        {xie2024bridging}
\bibfield{author}{\bibinfo{person}{Wenjia Xie}, \bibinfo{person}{Rui Zhou}, \bibinfo{person}{Hao Wang}, \bibinfo{person}{Tingjia Shen}, {and} \bibinfo{person}{Enhong Chen}.} \bibinfo{year}{2024}\natexlab{b}.
\newblock \showarticletitle{Bridging User Dynamics: Transforming Sequential Recommendations with Schr{\"o}dinger Bridge and Diffusion Models}. In \bibinfo{booktitle}{\emph{Proceedings of the 33rd ACM International Conference on Information and Knowledge Management}}. \bibinfo{pages}{2618--2628}.
\newblock


\bibitem[Xu et~al\mbox{.}(2024)]%
        {xu2024multi}
\bibfield{author}{\bibinfo{person}{Xiang Xu}, \bibinfo{person}{Hao Wang}, \bibinfo{person}{Wei Guo}, \bibinfo{person}{Luankang Zhang}, \bibinfo{person}{Wanshan Yang}, \bibinfo{person}{Runlong Yu}, \bibinfo{person}{Yong Liu}, \bibinfo{person}{Defu Lian}, {and} \bibinfo{person}{Enhong Chen}.} \bibinfo{year}{2024}\natexlab{}.
\newblock \showarticletitle{Multi-granularity Interest Retrieval and Refinement Network for Long-Term User Behavior Modeling in CTR Prediction}.
\newblock \bibinfo{journal}{\emph{arXiv preprint arXiv:2411.15005}} (\bibinfo{year}{2024}).
\newblock


\bibitem[Yang et~al\mbox{.}(2020)]%
        {ONN}
\bibfield{author}{\bibinfo{person}{Yi Yang}, \bibinfo{person}{Baile Xu}, \bibinfo{person}{Shaofeng Shen}, \bibinfo{person}{Furao Shen}, {and} \bibinfo{person}{Jian Zhao}.} \bibinfo{year}{2020}\natexlab{}.
\newblock \showarticletitle{Operation-aware neural networks for user response prediction}.
\newblock \bibinfo{journal}{\emph{Neural Networks}}  \bibinfo{volume}{121} (\bibinfo{year}{2020}), \bibinfo{pages}{161--168}.
\newblock


\bibitem[Ye et~al\mbox{.}(2025)]%
        {ye2025fuxi}
\bibfield{author}{\bibinfo{person}{Yufei Ye}, \bibinfo{person}{Wei Guo}, \bibinfo{person}{Jin~Yao Chin}, \bibinfo{person}{Hao Wang}, \bibinfo{person}{Hong Zhu}, \bibinfo{person}{Xi Lin}, \bibinfo{person}{Yuyang Ye}, \bibinfo{person}{Yong Liu}, \bibinfo{person}{Ruiming Tang}, \bibinfo{person}{Defu Lian}, {et~al\mbox{.}}} \bibinfo{year}{2025}\natexlab{}.
\newblock \showarticletitle{FuXi-$\alpha $: Scaling Recommendation Model with Feature Interaction Enhanced Transformer}.
\newblock \bibinfo{journal}{\emph{arXiv preprint arXiv:2502.03036}} (\bibinfo{year}{2025}).
\newblock


\bibitem[Yin et~al\mbox{.}(2024a)]%
        {yin2024learning}
\bibfield{author}{\bibinfo{person}{Mingjia Yin}, \bibinfo{person}{Hao Wang}, \bibinfo{person}{Wei Guo}, \bibinfo{person}{Yong Liu}, \bibinfo{person}{Zhi Li}, \bibinfo{person}{Sirui Zhao}, \bibinfo{person}{Zhen Wang}, \bibinfo{person}{Defu Lian}, {and} \bibinfo{person}{Enhong Chen}.} \bibinfo{year}{2024}\natexlab{a}.
\newblock \showarticletitle{Learning partially aligned item representation for cross-domain sequential recommendation}.
\newblock \bibinfo{journal}{\emph{arXiv preprint arXiv:2405.12473}} (\bibinfo{year}{2024}).
\newblock


\bibitem[Yin et~al\mbox{.}(2024b)]%
        {yin2024dataset}
\bibfield{author}{\bibinfo{person}{Mingjia Yin}, \bibinfo{person}{Hao Wang}, \bibinfo{person}{Wei Guo}, \bibinfo{person}{Yong Liu}, \bibinfo{person}{Suojuan Zhang}, \bibinfo{person}{Sirui Zhao}, \bibinfo{person}{Defu Lian}, {and} \bibinfo{person}{Enhong Chen}.} \bibinfo{year}{2024}\natexlab{b}.
\newblock \showarticletitle{Dataset regeneration for sequential recommendation}. In \bibinfo{booktitle}{\emph{Proceedings of the 30th ACM SIGKDD Conference on Knowledge Discovery and Data Mining}}. \bibinfo{pages}{3954--3965}.
\newblock


\bibitem[Yin et~al\mbox{.}(2023)]%
        {yin2023apgl4sr}
\bibfield{author}{\bibinfo{person}{Mingjia Yin}, \bibinfo{person}{Hao Wang}, \bibinfo{person}{Xiang Xu}, \bibinfo{person}{Likang Wu}, \bibinfo{person}{Sirui Zhao}, \bibinfo{person}{Wei Guo}, \bibinfo{person}{Yong Liu}, \bibinfo{person}{Ruiming Tang}, \bibinfo{person}{Defu Lian}, {and} \bibinfo{person}{Enhong Chen}.} \bibinfo{year}{2023}\natexlab{}.
\newblock \showarticletitle{Apgl4sr: A generic framework with adaptive and personalized global collaborative information in sequential recommendation}. In \bibinfo{booktitle}{\emph{Proceedings of the 32nd ACM international conference on information and knowledge management}}. \bibinfo{pages}{3009--3019}.
\newblock


\bibitem[Yin et~al\mbox{.}(2024c)]%
        {yin2024entropy}
\bibfield{author}{\bibinfo{person}{Mingjia Yin}, \bibinfo{person}{Chuhan Wu}, \bibinfo{person}{Yufei Wang}, \bibinfo{person}{Hao Wang}, \bibinfo{person}{Wei Guo}, \bibinfo{person}{Yasheng Wang}, \bibinfo{person}{Yong Liu}, \bibinfo{person}{Ruiming Tang}, \bibinfo{person}{Defu Lian}, {and} \bibinfo{person}{Enhong Chen}.} \bibinfo{year}{2024}\natexlab{c}.
\newblock \showarticletitle{Entropy law: The story behind data compression and llm performance}.
\newblock \bibinfo{journal}{\emph{arXiv preprint arXiv:2407.06645}} (\bibinfo{year}{2024}).
\newblock


\bibitem[Zhang et~al\mbox{.}(2024a)]%
        {Wukong}
\bibfield{author}{\bibinfo{person}{Buyun Zhang}, \bibinfo{person}{Liang Luo}, \bibinfo{person}{Yuxin Chen}, \bibinfo{person}{Jade Nie}, \bibinfo{person}{Xi Liu}, \bibinfo{person}{Daifeng Guo}, \bibinfo{person}{Yanli Zhao}, \bibinfo{person}{Shen Li}, \bibinfo{person}{Yuchen Hao}, \bibinfo{person}{Yantao Yao}, {et~al\mbox{.}}} \bibinfo{year}{2024}\natexlab{a}.
\newblock \showarticletitle{Wukong: Towards a Scaling Law for Large-Scale Recommendation}.
\newblock \bibinfo{journal}{\emph{arXiv preprint arXiv:2403.02545}} (\bibinfo{year}{2024}).
\newblock


\bibitem[Zhang et~al\mbox{.}(2019)]%
        {FAT-DeepFFM}
\bibfield{author}{\bibinfo{person}{Junlin Zhang}, \bibinfo{person}{Tongwen Huang}, {and} \bibinfo{person}{Zhiqi Zhang}.} \bibinfo{year}{2019}\natexlab{}.
\newblock \showarticletitle{FAT-DeepFFM: Field attentive deep field-aware factorization machine}.
\newblock \bibinfo{journal}{\emph{arXiv preprint arXiv:1905.06336}} (\bibinfo{year}{2019}).
\newblock


\bibitem[Zhang et~al\mbox{.}(2025b)]%
        {zhang2025td3}
\bibfield{author}{\bibinfo{person}{Jiaqing Zhang}, \bibinfo{person}{Mingjia Yin}, \bibinfo{person}{Hao Wang}, \bibinfo{person}{Yawen Li}, \bibinfo{person}{Yuyang Ye}, \bibinfo{person}{Xingyu Lou}, \bibinfo{person}{Junping Du}, {and} \bibinfo{person}{Enhong Chen}.} \bibinfo{year}{2025}\natexlab{b}.
\newblock \showarticletitle{TD3: Tucker Decomposition Based Dataset Distillation Method for Sequential Recommendation}.
\newblock \bibinfo{journal}{\emph{arXiv preprint arXiv:2502.02854}} (\bibinfo{year}{2025}).
\newblock


\bibitem[Zhang et~al\mbox{.}(2024b)]%
        {FusionChallenge2}
\bibfield{author}{\bibinfo{person}{Kexin Zhang}, \bibinfo{person}{Fuyuan Lyu}, \bibinfo{person}{Xing Tang}, \bibinfo{person}{Dugang Liu}, \bibinfo{person}{Chen Ma}, \bibinfo{person}{Kaize Ding}, \bibinfo{person}{Xiuqiang He}, {and} \bibinfo{person}{Xue Liu}.} \bibinfo{year}{2024}\natexlab{b}.
\newblock \showarticletitle{Fusion Matters: Learning Fusion in Deep Click-through Rate Prediction Models}.
\newblock \bibinfo{journal}{\emph{arXiv preprint arXiv:2411.15731}} (\bibinfo{year}{2024}).
\newblock


\bibitem[Zhang et~al\mbox{.}(2025a)]%
        {zhang2025killingbirdsstoneunifying}
\bibfield{author}{\bibinfo{person}{Luankang Zhang}, \bibinfo{person}{Kenan Song}, \bibinfo{person}{Yi~Quan Lee}, \bibinfo{person}{Wei Guo}, \bibinfo{person}{Hao Wang}, \bibinfo{person}{Yawen Li}, \bibinfo{person}{Huifeng Guo}, \bibinfo{person}{Yong Liu}, \bibinfo{person}{Defu Lian}, {and} \bibinfo{person}{Enhong Chen}.} \bibinfo{year}{2025}\natexlab{a}.
\newblock \bibinfo{title}{Killing Two Birds with One Stone: Unifying Retrieval and Ranking with a Single Generative Recommendation Model}.
\newblock
\newblock
\showeprint[arxiv]{2504.16454}~[cs.IR]
\urldef\tempurl%
\url{https://arxiv.org/abs/2504.16454}
\showURL{%
\tempurl}


\bibitem[Zhang et~al\mbox{.}(2024c)]%
        {zhang2024unified}
\bibfield{author}{\bibinfo{person}{Luankang Zhang}, \bibinfo{person}{Hao Wang}, \bibinfo{person}{Suojuan Zhang}, \bibinfo{person}{Mingjia Yin}, \bibinfo{person}{Yongqiang Han}, \bibinfo{person}{Jiaqing Zhang}, \bibinfo{person}{Defu Lian}, {and} \bibinfo{person}{Enhong Chen}.} \bibinfo{year}{2024}\natexlab{c}.
\newblock \showarticletitle{A Unified Framework for Adaptive Representation Enhancement and Inversed Learning in Cross-Domain Recommendation}. In \bibinfo{booktitle}{\emph{International Conference on Database Systems for Advanced Applications}}. Springer, \bibinfo{pages}{115--130}.
\newblock


\bibitem[Zhou et~al\mbox{.}(2019)]%
        {DIEN}
\bibfield{author}{\bibinfo{person}{Guorui Zhou}, \bibinfo{person}{Na Mou}, \bibinfo{person}{Ying Fan}, \bibinfo{person}{Qi Pi}, \bibinfo{person}{Weijie Bian}, \bibinfo{person}{Chang Zhou}, \bibinfo{person}{Xiaoqiang Zhu}, {and} \bibinfo{person}{Kun Gai}.} \bibinfo{year}{2019}\natexlab{}.
\newblock \showarticletitle{Deep interest evolution network for click-through rate prediction}. In \bibinfo{booktitle}{\emph{Proceedings of the AAAI conference on artificial intelligence}}, Vol.~\bibinfo{volume}{33}. \bibinfo{pages}{5941--5948}.
\newblock


\bibitem[Zhou et~al\mbox{.}(2018)]%
        {DIN}
\bibfield{author}{\bibinfo{person}{Guorui Zhou}, \bibinfo{person}{Chengru Song}, \bibinfo{person}{Xiaoqiang Zhu}, \bibinfo{person}{Ying Fan}, \bibinfo{person}{Han Zhu}, \bibinfo{person}{Xiao Ma}, \bibinfo{person}{Yanghui Yan}, \bibinfo{person}{Junqi Jin}, \bibinfo{person}{Han Li}, {and} \bibinfo{person}{Kun Gai}.} \bibinfo{year}{2018}\natexlab{}.
\newblock \bibinfo{title}{Deep Interest Network for Click-Through Rate Prediction}.
\newblock
\newblock
\showeprint[arxiv]{1706.06978}~[stat.ML]
\urldef\tempurl%
\url{https://arxiv.org/abs/1706.06978}
\showURL{%
\tempurl}


\bibitem[Zhu et~al\mbox{.}(2018)]%
        {TDM}
\bibfield{author}{\bibinfo{person}{Han Zhu}, \bibinfo{person}{Xiang Li}, \bibinfo{person}{Pengye Zhang}, \bibinfo{person}{Guozheng Li}, \bibinfo{person}{Jie He}, \bibinfo{person}{Han Li}, {and} \bibinfo{person}{Kun Gai}.} \bibinfo{year}{2018}\natexlab{}.
\newblock \showarticletitle{Learning tree-based deep model for recommender systems}. In \bibinfo{booktitle}{\emph{Proceedings of the 24th ACM SIGKDD international conference on knowledge discovery \& data mining}}. \bibinfo{pages}{1079--1088}.
\newblock


\bibitem[Zhu et~al\mbox{.}(2023)]%
        {FinalNet}
\bibfield{author}{\bibinfo{person}{Jieming Zhu}, \bibinfo{person}{Qinglin Jia}, \bibinfo{person}{Guohao Cai}, \bibinfo{person}{Quanyu Dai}, \bibinfo{person}{Jingjie Li}, \bibinfo{person}{Zhenhua Dong}, \bibinfo{person}{Ruiming Tang}, {and} \bibinfo{person}{Rui Zhang}.} \bibinfo{year}{2023}\natexlab{}.
\newblock \showarticletitle{Final: Factorized interaction layer for ctr prediction}. In \bibinfo{booktitle}{\emph{Proceedings of the 46th International ACM SIGIR Conference on Research and Development in Information Retrieval}}. \bibinfo{pages}{2006--2010}.
\newblock


\end{thebibliography}




\end{document}